\documentclass{aa}
\usepackage{graphicx}
\usepackage{txfonts}
\usepackage{rotating}
\usepackage{natbib}
\bibpunct{(}{)}{;}{a}{}{,} 

\newcommand{\reff}{\ensuremath{r_{\mathrm{e}}}}
\newcommand{\rmin}{\ensuremath{r_{\mathrm{min}}}}
\newcommand{\nser}{\ensuremath{n_{\mathrm{Ser}}}}

\begin{document}

\title{HST/ACS observations of shell galaxies: inner shells, shell colours and dust.
	\thanks{Based on observations made with the NASA/ESA Hubble Space Telescope, obtained at the Space Telescope Science Institute, which is operated by the Association of Universities for Research in Astronomy, Inc., under NASA contract NAS 5-26555. These observations are associated with program GO9399 and GO9427}}

\author{G.Sikkema\inst{1} \and  D.Carter\inst{2} R.F. Peletier\inst{1} \and M. Balcells\inst{3} \and C. del Burgo\inst{4} \and E.A. Valentijn\inst{1} }

\institute{Kapteyn Astronomical Institute, University of Groningen, PO Box 800, 9700 AV Groningen, The Netherlands  \and Astrophysics Research Institute, Liverpool John Moores University, 12 Quays House, Egerton Wharf, Birkenhead, CH41 1LD, United Kingdom \and Instituto de Astrof\'isica de Canarias, V\'ia L\'actea s/n, La Laguna, 38200, Spain \and School of Cosmic Physics, Dublin Institute for Advanced Studies,
Dublin 2, Ireland}

\date{}

\abstract
{Shells in Elliptical Galaxies are faint, sharp-edged features, believed to provide evidence for a merger event. Accurate photometry at high spatial resolution is needed to learn on presence of inner shells, population properties of shells, and dust in shell galaxies.
}
{Learn more about the origin of shells and dust in early type galaxies.
}
{V-I colours of shells and underlying galaxies are derived, using HST Advanced Camera for Surveys (ACS) data. A galaxy model is made locally in wedges and subtracted to determine shell profiles and colours. We applied Voronoi binning to our data to get smoothed colour maps of the galaxies. Comparison with N-body simulations from the literature gives more insight to the origin of the shell features. Shell positions and dust characteristics are inferred from model galaxy subtracted images.
}
{The ACS images reveal shells well within the effective radius in some galaxies (at 0.24$\,\reff$ = 1.7 kpc in the case of NGC~5982).  In some cases, strong nuclear dust patches prevent detection of inner shells. Most shells have colours which are similar to the underlying galaxy. Some inner shells are redder than the galaxy. All six shell galaxies show out of dynamical equilibrium dust features, like lanes or patches, in their central regions. Our detection rate for dust in the shell ellipticals is greater than that found from HST archive data for a sample of normal early-type galaxies, at the 95\% confidence level.                  
}
{The merger model describes better the shell distributions and morphologies than the interaction model. Red shell colours are most likely due to the presence of dust and/or older stellar populations. The high prevalence and out of dynamical equilibrium morphologies of the central dust features point towards external influences being responsible for visible dust features in early type shell galaxies. Inner shells are able to manifest themselves in relatively old shell systems.}


\maketitle

\newpage
\ \\


\section{Introduction}
\label{sec:introduction}

Shell galaxies \citep{malin} have long been recognised as useful laboratories for learning on both the formation processes and the internal structure of elliptical galaxies.  Soon after their discovery, shells were identified as tracers of ''the splatter produced by a merger'' \citep{schweizer80}, more specifically minor mergers.  Hence shell galaxies provided candidate configurations to investigate details of the accretion process, such as the nature of the accreted galaxy, the dominant types of accretion orbits, the radial distribution of the accreted matter, and the connection of accretion events to AGN activity.  If shells trace specific orbit configurations of accreted stars, then they potentially contain  information on the three-dimensional shape of the galaxian potential.

\citet{prieur90} recognised different morphological categories of shell galaxies. Type I shell galaxies have shells antisymmetrically (interleaved) aligned along the major axis. Type II shells are placed all around the galaxy. Type III shells show both or irregular features. Numerical simulation work in the eighties and early nineties provided the most widely accepted framework for interpreting these shell morphologies in terms of mergers.   Quinn's phase-wrapping formalism provides an elegant explanation for Type I, interleaved shells.  Phase wrapping  recognises the discrete distribution of turning points for radially-injected stars that oscillate back and forth in the galaxian potential \citep{quinn}; the shells themselves are the loci of the turning points, where pile-up leads to increased surface brightness; these loci move outward in the galaxy as a consequence of the proportionality between orbital period and apocentre distance.  More complex shell systems of Types II and III can be produced by minor mergers from non-radial orbits, on non-spherical parent galaxies, or due to internal rotation in the accreted galaxy \citep{h0,h2}.  Shells may also result from ''space wrapping'' \citep{dup} in the absence of radial orbit turn-around, when line-of-sight integration leads to an increase in surface brightness for debris of a satellite accreted on a high-angular momentum orbit.  Finally, shells may result from major mergers between two disk galaxies, as a result of the return of tidal tail material \citep{h3}, whenever the bulge-to-disk ratio of the parent galaxies is low \citep{gonzalez05}.  

Models for shell formation not based on mergers have been proposed as well.  These include tidal interactions \citep{thom1,thom2} or asymmetric local star formation \citep{low}.  Several observational diagnostics may be used to test the various theories.  Shells are mostly observed in isolated environments; this may indicate either a lower formation rate or a shorter lifetime in denser environments; it may indicate younger ages for shell systems in empty environments  \citep{col}.  \citet{ft} noted that almost all galaxies which contain a kinematically decoupled core (KDC), also show shells, suggesting a relation between these galaxy properties.  These two properties give support to the merger/accretion model for shells.




How close to galaxy centres do shells exist is a matter of current interest.  Because shell detection requires some form of unsharp filtering, whether digital or photographic \citep{malin}, and that works best when brightness gradients are not pronounced, shells have more often been detected in the outer parts of galaxies.  However, inner shells contain useful information on the shell-making process.  The existence of inner shells requires that orbital energy and angular momentum be removed from the accreted stars before these are released into the potential of the larger galaxy, therefore inner shells imply that dynamical friction operated, and that the accreted galaxy disrupted late into the accretion process. Because shells trace accreted material with orbit apocentres at the shell radii, inner shells would provide strong evidence for the late build-up of the inner regions of ellipticals through accretion of small galaxies.  

To improve on our ability to detect inner shells, we used the ACS camera on board HST to image the inner parts of six well-known shell galaxies (see Table~\ref{info}).  The spatial sampling of the ACS, six times better than typical ground-based cameras, coupled to the absence of atmospheric blurring, provides for a more accurate modelling of the underlying galaxy brightness distribution, and a more accurate mapping of the shell profiles themselves.  In this paper we present photometric data in $V$ and $I$ for the galaxies and the detected shells.  

The HST images also allow a precise determination of shell colours.  The latter may provide useful diagnostics on the various shell formation models.  The interaction model predicts similar colours for the shells as the host galaxy, whereas significant differences in shell colours are possible in the merger models.  To date, observations give a confusing picture on shell colours.  Examples are found of shells that are redder; similar, or bluer, than the underlying galaxy.   In some cases, different authors report opposite colour differences (shell minus galaxy) for the same shell; we mention specific instances of this in Sect~\ref{sec:results:colours}.  Colour even seems to change \textsl{along} some shells; examples are NGC~2865 \citep{fort}, NGC~474 \citep{prieur90}, and NGC~3656 \citep{bal97}.   
Errors in shell colours are very sensitive to the correct modelling of the underlying light distribution.  HST images allow for a detailed modelling of the galaxy light distribution, especially near the centres, and should provide increased accuracy in the determination of shell colours.


Another important issue is the properties of the visible dust in the centres shell galaxies, which might say something about the dust visible in the centres early type galaxies in general. Our observations may help to learn more about dust origin and formation theories \citep{lauer}.



The paper presents a photometric analysis of the shells in the six shell galaxies imaged in our HST/ACS program.  The systems observed contain three type I galaxies (NGC~1344, NGC~3923 and NGC~5982) and two type II (NGC~474, NGC~2865) and one type III (NGC~7626).  
The HST/ACS images are analysed by applying the technique of Voronoi binning \citep{capel} on the single passband images, which yields high-S/N brightness and colour maps to see if the shells influence local colour gradients.. In an companion paper, the properties of the globular clusters were analysed \citep[][hereafter Paper~I]{sikkema}.

The data reduction is briefly summarised in Section 2. In Section 3 we describe how we obtained global parameters, the shell fluxes and colours, the production of colour maps and dust properties. In Section 4, we compare our observations with N-body simulations of shell galaxies using different models, discuss implications of shell colours and analyse the dust properties of shell galaxies. We summarise our main conclusions in the last Section.



\section {Observations and Data Reduction}


\begin{table*}
\centering
\begin{tabular} {l c c c r r r r r r r| r r r r}
\hline
\hline

Galaxy&	type &RA (J2000)&  DEC(J2000) & $A_V$ & $A_I$ & type & $m_V$  & $M_V$& $\sigma$ & \reff & $sky_V$ & $sky_I$ &  \nser & r  \\
\hline
(1) &    (2)    &   (3)    &    (4)   &   (5)     &   (6)  &   (7)     &  (8)  &    (9)     &  (10)&   (11)   &   (12)   & (13) & (14) & (15)\\
\hline
\object{NGC 474}&  II  &$1^h 20^m 06^s.7$  & $+03^\circ 24' 55''$ & 0.11& 0.07&E/S0  & 11.39 &-21.17 & 164 & 50 &   203 & 129 &$^*7.9$ & 1.0\\
\object{NGC 1344}& I &$3^h 28^m 19^s.7$  & $-31^\circ 04' 05''$ & 0.06& 0.04&E5  & 10.41 &-21.07 & 187 &  13 &     94 &  47 &$4.6$     & 1.9\\
\object{NGC 2865}& II  &$9^h 23^m30^s.2$   & $-23^\circ 09' 41''$ & 0.27& 0.16&E3-4& 11.30 &-21.59 & 230 &  27&   146 & 109 &$^*6.3$   & 1.2\\
\object{NGC 3923}& I &$11^h 51^m01^s.8$  & $-28^\circ 48' 22''$ & 0.27& 0.16&E4-5& 9.88  &-21.92 & 249 &   39&    170 & 103 &$5.1$     & 1.9\\
\object{NGC 5982}& I &$15^h 38^m39^s.8$  & $+59^\circ 21' 21''$ & 0.06& 0.04&E3  & 11.20 &-21.91 & 240 &   34&    124 & 182 &$5.1$     & 1.0\\
\object{NGC 7626}& III  &$23^h 20^m 42^s.3$ & $+08^\circ 13' 02''$ & 0.24& 0.14&Epec& 11.25 &-22.16 & 270 & 24&   119 &  80 &$7.0$     & 1.1\\
\hline
\end{tabular}
\caption{Properties of six shell galaxies. Columns (1-11) give data from the literature (with columns (1-10) from paper I and column (11): \citealp{vau}); columns (12-14) present data derived in this paper: (1): Galaxy name, (2): shell galaxy type, (3-4): Right Ascension and Declination in degrees, (5-6): Galactic extinction in V and I (in magnitudes), (7-9): Morphological type, apparent V band magnitude and absolute V band magnitude. (10): velocity dispersion $\sigma$ in km/s, (11): effective radius in arcsec. (12-13): Calculated background values of the ACS V and I images in counts. (14-15): fitted S\'ersic index n, starting from radius r (arcsec). *: no stable fit possible. Fits are drawn in Figures A.3-F.3.}
\label{info}
\end{table*}

The six shell galaxies were observed with the ACS\_WFC camera between July 2002 and January 2003 with the filters F606W (V-band) and F814W (I band) in CR\_SPLIT=2 mode . The camera contains two CCDs of 2048 x 4096 pixels, each pixel having a size of $0''.049$ pixel$^{-1}$ resulting in a field of view of $202''$ x $202''$. Exposure times were on average 1000s. The inner 24 pixels of NGC 2865 and the inner 8 pixels of NGC 474 were saturated in both V and I. Table 1 contains the main characteristics of the galaxies like Right Ascension and Declination at Epoch J2000.0, the extinction coefficients, exposure times and adopted distances throughout this paper. Detailed information about the data reduction can be found in Paper I.

In addition we found three B band (filter F435W) observations from July 2003 of NGC 7626, associated with program GO9427 in the HST archive with exposure times of 2620, 2620 and 2480 seconds.. We used the standard reduced images and combined them to remove most cosmic rays. We used a galactic extinction value of 0.313 \citep{schlegel} and applied the following transformation formulae \citep{sirianni}:

\begin{equation}
B_{J}=m(F435W)+25.709+0.108(B-V)_{JC}-0.068(B-V)_{JC}^2
\end{equation}
\begin{equation}
V_{J}=m(F606W)+26.410+0.170(B-V)_{JC}+0.060(B-V)_{JC}^2
\end{equation}

Here =m(F435W) and m(F606W) are the ACS instrumental magnitudes and B and V are in the Johnson-Cousins system. 

\section{Data Analysis}
\subsection{Global parameters}
\label{sec:globalparameters}

Information about the morphology of the galaxies and location of possible shells was obtained by using the ellipse fitting task GALPHOT (see \citealp{jorg}); it returns information such as ellipticity, position angle, surface brightness and the C3,C4,S3,S4 coefficients \citep{carter78}, all as a function of radius. A model galaxy subtracted residual image is returned as well, which is shown in Figures A.1 to F.1 for each galaxy.

In the GALPHOT processing, background galaxies, point-like objects, dust features, bright pronounced shells and additional bad data were masked out by hand in an iterative way. Remaining faint shell structures, having a brightness typically not more than 5\% of the underlying galaxy emission, do not notably affect the results. The best fits were obtained by allowing the centre, position angle and ellipticity as free parameters to vary. In two cases: NGC 2865 and NGC 5982 the central regions do not have reliable fits (see Figures C.2 and E.2 respectively). These regions correspond to rough circles with diameters $24''$ for NGC 5982 and $4''$ for NGC 2865 respectively. The pixels within $0.3''$ of the centre of the latter galaxy are saturated.

Global surface brightness profiles were obtained by plotting the surface brightness for each fitted ellipse as function of radius. The outer parts of the profiles are severely influenced by the background uncertainty. It is difficult to determine reliable background values from the ACS images themselves, since the galaxies fill the whole field. Fortunately, for all galaxies, apart from NGC 5982, we found optical wide field data in the R band in the ESO archive from the Wide Field Imager (WFI) at the 2.2m ESO/MPI telescope. The WFI camera has a field of view of $34'$ x $33'$, much larger than the galaxies. 

We used the ASTRO-WISE system \footnote{www.astro-wise.org/portal} \citep{vk} to reduce the WFI images. After subtracting a constant background value from the WFI images, GALPHOT was applied on them. 
The ACS background values were determined by matching ACS surface brightness profile to the WFI surface brightness profiles. For the galaxy without WFI data, NGC 5982, we assumed that the surface brightness profile follows a straight line from a certain point in a r$^{1/4}$-log(I) plot. The calculated values for the background are listed in Table 1.
The final GALPHOT results describing the morphology of the six galaxies are shown in Figures A3-F3 in Appendix A to F. These Figures also include the results of the isophotal analysis on P.A. and ellipticity from WFI ground based images. The errors in V-I (top-right panels) were determined using the scatter in the WFI backgrounds. We fitted S\'ersic profiles to the I band surface brightness data using of the equation:

\begin{equation}
\mu(r) = \mu_e + c_n \left[\left(\frac{r}{r_c}\right)^{1/\nser}-1\right]
\end{equation}

with $c_n = 2.5 (0.868n - 0.142)$  (valid for $ 0.5 < n < 16.5$; \citealp{caon}). The fit was made from that particular radius, chosen by eye, where the, sometimes visible, inner plateau, will not disturb the fit. These starting radii are drawn in Figures A-F.3 as vertical dashed lines. For two galaxies, NGC 474 and NGC 2865, no stable fits were possible, with large variations for $n$ depending on the starting point fitting radius r. The fitted values for n and starting points are listed in column (13-14) of Table 1.

For NGC~474, adding an outer exponential to the fitting function significantly improves the surface brightness profile fit, resulting in a smaller and less concentrated spheroid than that listed in Table~\ref{info}: $\reff = 6.6\arcsec$ and $\nser = 3.0$, and $B/D=0.71$.  Such 'bulge-disk' decomposition of the profile led \citet{schom} to argue that NGC~474 is a face-on S0, a point which we address in Sect.~\ref{sec:ngc474}.

Residual images are obtained after subtracting the galaxy models obtained by GALPHOT and are shown in Figures A1-F1. They were solely used to identify and locate the shell features, but not to determine the brightness of the shells: the residual images still show some large scale fluctuations in the background which will disturb measurements of faint shell fluxes significantly. This makes it difficult to obtain reliable shell brightness from these images. A better approach, described in the next Section, is to work locally, within wedges.

Except for NGC 1344, global isophotal analysis of our galaxies with other data has been done before: NGC 474 (ground-based B and V: \citealp{pier}; ground based B and R: \citealp{turn}), NGC 2865 (ground based B, V and I: \citealp{reid}), NGC 3923 (ground based B and R: \citealp{jedr}), NGC 3923, NGC 5982 and 7626 (ground based V, R and I: \citealp{bender}) and NGC 5982 and NGC 7626 (HST WFPC2, V and I: \citealp{car}; NICMOS 1.6 $\mu$m: \citealp{quillen}). As mentioned before we can also compare with WFI archive data for NGC 474, 1344, 2865, 3923 and 7626. Comparison with these data give similar results. 

\subsection{Shell radii}
\label{sec:analysis:shellradii}

Shell positions were determined by two of us (DC and GS) by visual inspection on an image display of the residual images described in \S~\ref{sec:globalparameters}.  Shell positions are listed in Table 2.  Following Prieur, we list radii corresponding to the outermost edge of each shell. 

Shell radii are discussed in \S~\ref{sec:results:shellradii}.
\begin{table*}
\label{tab:shellradii}
\centering
\begin{tabular} {c c c c c c c c c c}
\hline
\hline

Galaxy&	Direction&  $ a $ ('')& $\times$ \reff  &Comments & Galaxy&	Direction&  $ a $ ('')& $\times$ \reff  &Comments  \\ \cline{1-5}   \cline{6-10}  
\object{N 474}&     &  &   &    &   10$^*$                & S&  55.5&1.42&Prieur: 18S, 55.7"  \\  \cline{1-5}
1$^*$                &  N& 27.6&0.55&     &   11                & N&  60.4 &1.55&               \\
2                    &WSW& 31.8&0.64&     &   12                & N&  64.1 &1.64&               \\
3$^*$                &  S& 39.7&0.79& shells 3a and 3b   &   13$^*$                & S&  67.0 &1.72  & Prieur: 16S, 67.1"\\
4$^*$                &  W& 41.3&0.83&                    &   14                & N&  72.8 &1.87          &Prieur: 15N, 73.0"                         \\
5$^*$                &  W& 60.8&1.22&                    &   15$^*$                & S&  79.6 &2.04   &Prieur: 14S, 79.3"   \\
6$^*$                &  S& 61.1&1.22&                    &   16                & N&  99.9 &2.56    &                \\
7$^*$                &  N& 64.0&1.28& shells 7a and 7b, long arc &   17$^*$                & S& 103.6 &2.66   &Prieur: 12S, 104.7"  \\
8                &NNW& 74.9&1.50& diffuse    &   18                & N& 128.1 &3.28   &Prieur: 11N, 128.1" \\  \cline{6-10}
9$^*$                &  S& 76.2&1.52&                    &   \object{N 5982}&   &     &     &                 \\  \cline{6-10}
10                &  S& 77.2&1.54&     &   1                  & E&   8.0 &0.24   &       \\
11$^*$                &  W& 87.2&1.74&                  &   2                  & E&   9.8 &0.29   &     \\
12                & SW& 99.5&1.99& diffuse    &   3                  & W&  10.9 &0.32   &     \\
13$^*$                &  N&103.0&2.06& shells 13a and 13b, long arc     &   4                  & E&  12.5 &0.37   &     \\  \cline{1-5}
\object{N 1344}&   &     &   &   &   5                  & E&  15.0 &0.44   &     \\  \cline{1-5}
1$^*$            & NNW& 26.7&2.05 &      &   6                  & E&  17.7 &0.52   &     \\
2                & SSW& 37.0& 2.85&bright blob    &   7                  & E&  19.9 &0.59   &     \\
3$^*$            & NNW& 53.3&4.10 &      &   8                   & W&  20.1 &0.59   &     \\
4                & WSW& 57.8&4.45 &    &   9                   & E&  21.4 &0.63   &     \\
5                & NNW& 62.0&4.77 &diffuse    &   10                   & W&  21.9 &0.64   &     \\
6                & SSW& 65.5&5.04 &    &   11                   & E&  23.6 &0.69   &     \\
7                & WSW& 71.3&5.48 &    &   12                   & W&  23.9 &0.70   &     \\
8$^*$            & SSE& 93.1&7.16 &      &   13                   & E&  27.3 &0.80   &     \\
9                & SSE&109.5& 8.42&       &   14                   & W&  28.7 &0.84   &     \\
10$^*$           & SSE&122.6&     &      &   15                   & W&  31.9 &0.94   &     \\  \cline{1-5}
\object{N 2865}&   &        &   &   &   16                   & E&  39.0 &1.15   &     \\  \cline{1-5}
1                 &SW&   77.1 &2.86  &large     &   17                   & W&  39.5 &1.16   &     \\
2$^*$                 &E&    83.0 &3.07  & bright shell    &   18                   & W&  47.5 &1.40   &     \\
3                 &W-E&  90   &3.33  &large scale loop    &   19                  & E&  49.8 &1.46    &    \\
4                 &SE&   99.0 &3.67  &diffuse loop    &   20                   & E&  65.3 &1.92   &     \\  \cline{1-5}
\object{N 3923}&   &       &    &   &   21                   & W&  67.5 &1.99   &     \\  \cline{1-5}
1$^*$                & S&  18.0 &0.46&Prieur: 24S, 18.8"   &   22                   & W&  78.7 &2.31   &     \\
2                & N&  19.4 & 0.50 &Prieur: 23N, 19.5"             &   23                   & W&  91.8 &2.70   &\\
3$^*$                & S&  28.7 &0.74   &Prieur: 22S, 30.0"    &   24                   & NE& 100  &2.94   &          \\  \cline{6-10}
4                & N&  29.3 &0.75     &Prieur: 21N, 30.0"             &   \object{N 7626}&   &     &       &\\  \cline{6-10}
5                & N&  34.3 &0.88   &diffuse      &   1                & SW & 24.1&1.00    &\\
6                & S&  37.7 &0.97   &          &   2$^*$                & NE & 32.4&1.35   &  \\
7                & N&  41.5 &1.06   &                   &   3$^*$                & SW & 43.8&1.83   &  \\
8$^*$                & S&  44.0 &1.13   &Prieur: 20S. 44.7"     &   4                & E  & 47.6&1.98   &\\
9                & N&  51.2 &1.31 &                 & &  & & & \\
\hline \hline
\label{addshells2}
\end{tabular}
\caption{All shells identified by eye in our GALPHOT residual images. (1) Shell number/label. Asterisks indicate shells which have a colour determination (see also Table 3 and the appendix); (2) Shell direction relative to the centre; (3) Semi-major axis of outer shell border in arcseconds as measured from the galaxy centre; (4) Similar to column 3 but now in terms of effective radius; (5) Comments: additional information. The comments for NGC 3923 give Prieur (1988) labels + positions.}
\end{table*}

\subsection{Shell fluxes}
\label{sec:analysis:shellfluxes}

We have developed a three-step procedure to determine the shell brightness. Common to all the steps is that we work locally, within wedges, that are carefully placed over parts of the shells. For a particular galaxy the same wedges, shown in Figures A.1-F.1, were used in the two passbands. Next follows a detailed description of the procedure.

\begin{enumerate}
\item{Determining the local surface brightness profile of the galaxy.} \\
First the surface brightness profile within the wedge was derived. Data points of the curve were calculated by averaging the pixel values within the partial elliptical rings covered by the wedge. The elliptical ring segment had a thickness of 2.5 pixels, with a fixed central point and an ellipticity and position angle, whose values were taken by averaging the I-band GALPHOT results of the outer galaxy regions. Pixels within the wedge belonging to GALPHOT masks (see Section 3.1) were not used. Remaining pixel outliers in the elliptical ring segment were removed by iterating 10 times over the set of pixel values, each time applying a 4 $\sigma$ clipping method. The resulting surface brightness profile valid for the wedge is used in the next step.

\item{Making a local galaxy model.} \\
Several smooth curves were fitted to the surface brightness profile using Legendre polynomials of different degrees, applying the IRAF tool 'CURFIT'. Usually the shells are already visible as small bumps in the profile. Existence of shells is double checked by inspecting the residual images of GALPHOT. Another check is made by inspecting if the bumps are visible at the same position in both passbands. The data points in the profile which are part of the bumps in V and I, were left out in the fitting procedure. The fitted profile gives the model flux values in the partial elliptical rings mentioned in the previous step, which enables us to construct a 2d model image, which is valid locally for the wedge.

\item{Obtaining shell surface brightness.} \\
After subtracting each model image from the galaxy image, the residual images show the shells in each wedge for each Legendre model. The same recipe, as described in the first step, was applied again to obtain surface brightness profiles for each residual image, but now using the ellipticities of the shells, which, in case of type I shell galaxies (NGC 1344, NGC 3923 and NGC 5982), are much rounder than the underlying galaxy (this was already known, see e.g. Prieur, 1988). The surface brightness profiles now clearly show the shell fluxes and the background is close to zero. The results are shown in the top panels of the Figures (i.e. 4, 5, 6 etc.) in Appendix A-F. Reliable shell fluxes could not be obtained for left-side (North-East) of NGC 3923 and NGC 5982, because they are not well defined features, which is related to their low contrast and S/N w.r.t. the galaxy.
\end{enumerate}

\subsection{Shell colours}
By combining the results of two passbands for each Legendre fit, we obtain the average colours in the shell regions. Due to the different fits (i.e. each fit is constructed using a particular Legendre polynomial), the derived colours show variations which increase for fainter shells. Consequently, we use only those shell regions for which all fits showed a stable answer. These regions are  indicated by vertical lines in the top panels of Figures 4, 5, 6 etc. in Appendix A-F and typically contain data points with at least 10 counts. The final shell colour within a region is calculated by applying a weighted average, using the values for the colours and their errors derived for each Legendre fit. The resulting values are plotted in the Figures and listed in Table 3 .


\subsection{Galaxy colours}
Global colour maps of the galaxies were obtained by using the adaptive binning algorithm \citep{capel}. This algorithm bins two-dimensional data to a constant signal-to-noise ratio per bin by calculating a Voronoi tessellation. For objects with large gradients in S/N, as is the case for galaxies, this will result in smooth 2D colour images in regions with low S/N and therefore will show the colour gradients better than in the traditional way.
A practical example and explanation of this algorithm is given in \citet{ferreras}.

The procedure to obtain the colour maps consists of three steps:

First, the adaptive binning algorithm is applied on the 3x3 binned non-background subtracted I band images by using a S/N of 250 and leaving out masked regions. The resulting Voronoi tessellation is also used for the V band image in the next steps. The Voronoi tessellation is further processed by applying a Delaunay triangulation to the central positions of the Voronoi cells, assigning the average flux values within the cells to these central positions. Finally, V-I colour maps are obtained by subtracting the appropriate background values and combining the images by applying the appropriate transformation formulae (see Paper I). 

In the outer regions, where the uncertainties in the background become important, we still see large variations. These outer regions also show discrete offsets between different quadrants which increase when going outwards. This is due to random variations of the subtracted bias level as measured in the overscan versus the actual bias level in the science images\footnote{http://www.stsci.edu/hst/acs/documents/isrs/isr0407.pdf}. These differences can be as large as a few counts, which will show up especially in the V-I colour maps with low signal: the outer galaxy regions in our images. Unfortunately, appropriate calibration data to correct for this effect only exists for ACS observations later than November 2004, much later than our observations.

NGC 474 and NGC 3923 show elliptical red rings which are approximately 0.05 mag. higher in V-I than its surroundings (Figures 1 and 2 respectively). We believe that these are caused by artifacts, probably reflections, within the optical system. We double checked this for NGC 3923 by constructing a V-I colour map from ground based VLT-FORS2 images in Bessel V and I. The colour map does not show the ring which is illustrated in the lower panel of Figure D.4 showing, as a green line, the V-I profile of the ground based data: the bump due the ring between r=40'' and 60'' is not visible.


Any correlation between shells and integrated V-I colours is checked by calculating the Voronoi colours in the same wedges as used in the previous section; in general the shell fluxes are so low that little of it is reflected in the integrated V-I colour profiles. This can be seen in the lower parts in Figures 4, 5, 6 etc. in the Appendix A-F. We also see that most shell V-I colours are usually similar or sometimes redder than the colour of the galaxy. The signature of the rings of NGC 474 and NGC 3923 is also visible in these Figures as shallow, large scale, bumps with an amplitude about 0.05 at 25 arcsec and 50 arcsec for NGC 474 and NGC 3923 respectively. For NGC 7626, where we also have B data, we only could determine a reliable shell colour for shell 2, which has $(B-V)_{galaxy}=0.99\pm0.02$ and $(B-V)_{shell}=1.11\pm0.05$ (see Figure 6).

\begin{figure}
    \includegraphics[height=8cm,width=8cm, angle=0]{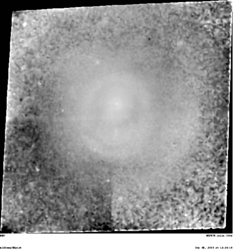}
    \caption{Voronoi binned colour image of NGC 474. The ring is probably an artifact; offsets between quadrants are also visible.}
    \label{colour474}
\end{figure}

\begin{figure}
    \includegraphics[height=8cm,width=8cm, angle=0]{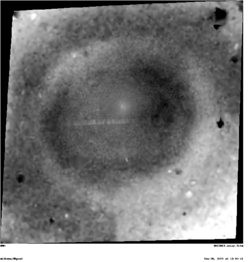}
    \caption{Similar as Figure 1 but now for NGC 3923.}
    \label{colour3923}
\end{figure}






\subsection{Shell radial profiles}
Figure 3 shows profiles for 21 bright shells in several galaxies. Some shells show plateaus and are asymmetric: they reach a maximum flux near the outer shell border, often accompanied with a sudden sharp drop (examples: N474\_6, N474\_13, N2865\_2, N3923\_8, N3923\_10, N7626\_2, N7626\_3). Other shells have a symmetrical Gaussian-like shape (N474\_7, N474\_11, N3923\_1, N3923\_15, N1344\_3, N1344\_6). Some shells seem to have double peaks (not shown in Figure but see the Appendix). A quantitative characterisation is obtained by applying a Gauss Hermite fitting procedure \citep{marel} using five free parameters: $\gamma$, $R_0$, $\sigma$, H3 and H4. The fitted values are given in Table 3 and the results are drawn as red curves in Figure 3.

The shells we see are the result of a projected 3D density distribution in our line of sight. To get more information about the real or intrinsic three dimensional density distribution, we integrated along the line of sight using two different simple 3D density shell models assuming spherical symmetry and assuming a opening angle $\phi$ in our line of sight. Two other parameters used in both models are the points $r_0$ where shells have their maximum stellar density $\rho_0$.

\begin{figure*}
    \includegraphics[width=17cm, angle=0]{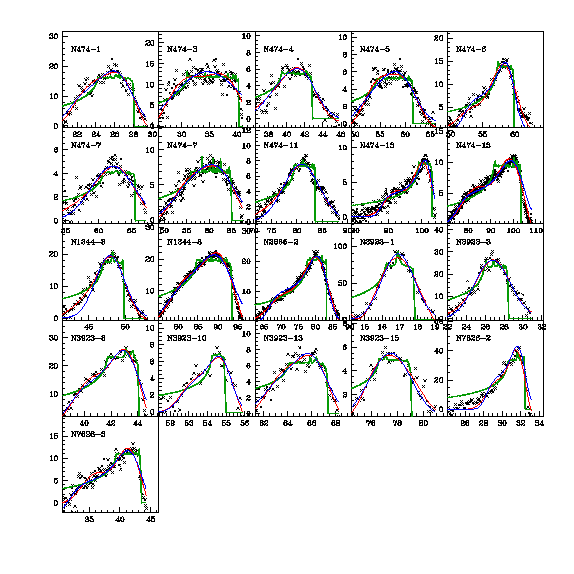}
    \caption{Shell profiles with different fits. Horizontal scale in arcsec. Vertical scale in counts. The red curve is a Gauss Hermite fit to the data giving a quantitative characterisation of the shell shapes. Green and blue curves are the results by integrating along our line of sight a spherical symmetrical density model using a $(r_0-r)^{-0.5}$ and Gaussian distribution respectively.} 
    \label{shellprofiles}
\end{figure*}

The first model assumes that the shell has an intrinsic  $(r_0-r)^{-0.5}$ density distribution which is predicted by theory to describe the inner parts of phase wrapped shells \citep{dup, prieur}. The second model is supposed to describe spatial wrapped shells \citep{turn_coll}. It assumes that shells are have an intrinsic Gaussian shape with thickness $r_g$. However this model lacks a physical basis. The results are again given in Table 3 and shown in Figure 3: here the green and blue curves depict the results of the first and second model respectively. In most cases the second model describes the data best. The 'noise' in the green curve is due to the fact that a discontinuous model (the model has a sudden drop in flux to zero at $r_0$) is fitted to the data: the noise in the model reflects the noise in the data.

\section{Notes on the individual systems}
\label{sec:individualsystems}

\subsection{NGC 474}
\label{sec:ngc474}

The low $v/\sigma$ ratio (this defines the amount of rotation), of 0.18 \citep{ramp} is consistent with NGC 474 being either a near face-on disk, or a near spherical galaxy. NGC 474 is part of a small evolving poor group \citep{ramp}. Its shell system is very complicated as shown in Figure 1 of the multi-wavelength study of Rampazzo et al.. In addition to the shells, a loop is visible, which heads east-west, starting from a comma shaped feature (outside the ACS field of view), passing the sideline towards the galaxy centre. In H{\sc I}, there are signs of tidal interactions with the nearby (at 5.5') regular spiral galaxy NGC 470. The same study does not make clear, however, if NGC 470 is responsible for or even related to the shell system. 

Being classified as a type II shell galaxy, this system has been used to test predictions of the weak interaction model (WIM; \citealp{thom1, thom2}). Here, shells are induced in the outer parts of the host galaxy due to tidal effects resulting from a fly-by of another galaxy. Morphologically, the WIM simulations show the shells as almost complete windings or spirals around the centre, when looking face-on. Looking at our GALPHOT residual image of NGC 474, however, the shells look more like short arcs. The shell positions and shapes (see the residual image Figure A.1) resemble more those of the results of the merger simulations shown in e.g. Figure 5 of \citet{dup}, model 4 of \citet{h2}, or model 7 of \citet{h0}: all mergers of low-mass companions on non-radial orbits with a spherical or mildly oblate primary. In these merger simulations the shells can be very old ($>$5Gyr) and are spatially wrapped around the galaxy.


Except for one shell, the colours of the analysed shells (see Table 3 and Figures A.4-A.8) are similar to the galaxy, consistent with previous studies (in B-V, V-r, R-I (Schombert \& Wallin, 1987), in B-R (Turnbull et al, 1999) and in B-V \citep{pier}). Some large shells are overlapped by different wedges, enabling us to compare their colours independently: shell 3 is located at $\approx$ 40'' from the centre. We splitted the shell into shells 3a and 3b (see Figure A.1); comparing their colours gives similar V-I colours: 1.21 $\pm0.13$ for 3a and  1.17$ \pm0.05$ for 3b. Similarly, shell 7 was splitted into shells 7a and 7b (located at about 64'' from the centre, see Figure A.1). They have V-I colours of 0.96 $ \pm0.05$ and 1.01 $ \pm0.08$ respectively. For shell 13a and 13b, at 103'' from the centre, we find significantly different colours: 0.93$ \pm0.04$ and 1.14$ \pm0.05$. These are probably erroneous values due to low shell fluxes compared to the underlying galaxy. \citet{wilk} also find offsets in colour up to 0.30 mag between shell segments a similar type II shell galaxy 0422-476. Another colour determination for this shell is given by \citet{schom}, who find redder R-I colours than the galaxy (galaxy R-I=0.88, shell R-I=1.09) and Pierfederici \& Rampazzo (2004) who find slightly redder colours in B-V.  The innermost sharp edged shell is detected at about 30'' from the centre in the South direction. 

Of all shells analysed in this work, the only really blue shell relative to the integrated galaxy colour is found in this galaxy, which is shell 5. Interestingly, \citep{turn}, found that their only really blue shell w.r.t. the integrated galaxy colour is the comma shaped feature at the SW (beyond the field of our ACS images, but see Figure 9 \citealp{turn}). We note that the position of our blue shell 5 lies exactly on the tail or loop, which is connected to the comma shaped feature. We therefore suggest that shell 5 is related to this feature. Blue shell colours in other shell galaxies have been found in young, gas-rich merger remnants such as NGC 3656 \citep{bal97} and Arp 230 (McGaugh \& Bothun, 1990) as well as blue tails in many other interactions (see e.g. \citealp{schombert2}). It is therefore tempting to conclude that this entire feature is the remains of a recent small merger unrelated to the rest of the red shell system.

Isophotal analysis of NGC 474 (Figure A.3) shows that the ellipticity is changing fast from 0.08 to 0.24 between 10'' and 20''. and back to 0.08 beyond 20''. At the same radii, the position angle is changing from 0 to about 20 and back. The galaxy contains several pronounced dust lanes within the inner 15'', not seen previously \citep[see, e.g.,][]{ravi,sarzi}.  \citeauthor{ravi} find a point source in the centre. The top right panel of Figure A.3 shows that this source is 0.05 mag. bluer in colour than its surroundings ($V-I=1.41$). NGC 474 contains the largest visible dust mass of our sample ($\approx 10^4 M_\odot$, Table 4) and its centre shows peculiar kinematic behaviour \citep{hau2}. 

\begin{sidewaystable*}
\centering
\begin{tabular} {c c c c | c c c c | c c c | c c c c c | c}
\hline
\hline

Galaxy&	$V-I_{shell}$ & $V-I_{gal}$ & $\Delta_{V-I}$ & $R_0$ & $\Phi$ & $\rho$ & $R_g$ &  $R_0$ & $\Phi$ & $\rho$ & $\gamma$ & $R_0$ & $\sigma$ & $h_3$&$h_4$& Comments\\
\hline
(1) &    (2)    &   (3)    &    (4)  & (5)  &(6)  &(7) &(8) &(9)  &(10)& (11)&(12)& (13)& (14)& (15) &(16) & (17)\\
\hline
\object{N 474}&     &     &  &  &    &    &  &   &  & &  & &  & &  &       P.A.=5.0 and ellipticity=0.13\\
\hline
1     & 1.46$ \pm0.06$  &  1.21 $\pm0.02$  &  +0.25 $\pm0.06$ & 27.1 & 39.1 & 2.0 & 1.8 & 28.0 & 30.5 & 1.5 & 78.5 & 57.6 & 2.4 & -0.2 & 0.1 & \\
3a    & 1.21$ \pm0.13$ &  1.21 $\pm0.02$  &   0.00 $\pm0.13$&   &  & & &  & &  & &  & &  & &\\
3b    &1.17$ \pm0.05$ & 1.21  $\pm0.02$  &  -0.04 $\pm0.05$ & 37.4 & 41.0 & 7.3 & 0.6 & 40.3 & 40.0 & 0.9 & 98.7 & 25.1 & 2.1 & -0.1 & -0.2 & \\
4     & 1.46$ \pm0.06$  &  1.20 $\pm0.02$  &  +0.26 $\pm0.06$ & 42.3 & 28.0 & 2.0 & 0.5 & 42.8 & 23.0 & 0.4 & 67.1 & 80.7 & 3.6 & -0.1 & 0.0 & \\
5     & 0.96$ \pm0.08$  &  1.15 $\pm0.02$  &  -0.19 $\pm0.08$ & 60.3 & 32.0 & 3.8 & 0.3 & 61.5 & 28.0 & 0.3 & 68.7 & 98.1 & 4.0 & -0.4 & -0.1 & \\
6     & 1.09$ \pm0.08$ &  1.06 $\pm0.02$  &  +0.03 $\pm0.08$ & 59.4 & 29.0 & 1.4 & 1.1 & 59.8 & 17.0 & 0.8 & 152.0 & 34.3 & 4.3 & -0.0 & -0.2 & \\
7a    & 0.96$ \pm0.05$  &  1.03 $\pm0.02$  &  -0.07 $\pm0.05$ & 62.5 & 12.5 & 4.3 & 0.4 & 65.5 & 22.3 & 0.2 & 30.3 & 40.7 & 2.0 & -0.1 & -0.0 & \\
7b    &1.01$ \pm0.08$ & 1.07 $\pm0.02$ & -0.06 $\pm0.08$&   &  & & &  & &  & &  & &  & &\\
11    & 1.42$ \pm0.05$  &  1.17 $\pm0.02$  &  +0.25 $\pm0.05$ & 80.9 & 7.0 & 4.8 & 0.8 & 83.5 & 18.0 & 0.4 & 192.6 & 93.4 & 8.0 & -0.3 & -0.2 & \\
13a   & 0.93$ \pm0.04$  &  1.01 $\pm0.02$  &  -0.08 $\pm0.04$ & 102.5 & 43.1 & 5.7 & 0.3 & 103.4 & 28.0 & 0.4 & 54.6 & 57.2 & 3.4 & -0.0 & -0.1 & \\
13b   &1.14$ \pm0.05$ & 1.01 $\pm0.02$ & +0.13 $\pm0.05$&   &  & & &  & &  & &  & &  & &\\
\hline
\object{N 1344}&   &     &   &  &   &   &  &   &  & &  & &  & &  &  P.A.=163.0 and ellipticity=0.38\\
\hline
1  & 1.39 $\pm0.11$& 1.23 $\pm0.01$ & +0.16 $\pm0.11$&  &  & & &  & &  & &  & &  & &\\
3  &  1.30 $\pm0.05$ &  1.20 $\pm0.01$  &  +0.10 $\pm0.05$ & 47.9 & 6.9 & 2.2 & 4.5 & 49.8 & 19.5 & 1.3 & 82.5 & 47.6 & 1.7 & -0.1 & 0.0 & \\
8  &  1.16 $\pm0.02$ &  1.21 $\pm0.01$  &  -0.05 $\pm0.02$ & 91.9 & 33.0 & 4.0 & 0.8 & 93.4 & 25.1 & 1.1 & 226.6 & 87.1 & 4.1 & -0.2 & -0.2 & \\
10 & 1.20 $\pm0.04$& 1.22 $\pm0.01$ & -0.02 $\pm0.04$&  &  & & &  & &  & &  & &  & &\\
\hline
\object{N 2865}&   &        &      &          &   &   &  & &  & &  & &                      &                & &      P.A.=152.0 and ellipticity=0.27\\
\hline
2, E  & 1.11 $\pm0.05$ &  0.98 $\pm0.01$ &  +0.13 $\pm0.02$ & 82.3 & 36.3 & 2.7 & 1.1 & 83.2 & 21.3 & 1.1 & 214.0 & 78.0 & 4.3 & -0.3 & -0.0 & Shell 2B of $\dagger$: $(V-R)_J=0.84 \pm0.09$   \\
\hline
\object{N 3923}&   &       &        &         &       &   &  & &  & &  & &                  &                 & &      P.A.=49.0 and ellipticity=0.35\\
\hline
1   &  1.27 $\pm0.03$  &  1.28 $\pm0.01$ &  -0.01 $\pm0.03$   & 17.4 & 23.5 & 0.9 & 18.0 & 17.9 & 25.2 & 8.9 & 324.9 & 14.0 & 1.8 & 1.0 & -1.2 & \\
3   &  1.37 $\pm0.05$  &  1.29 $\pm0.01$ &  +0.08 $\pm0.05$   & 27.3 & 21.0 & 2.1 & 3.5 & 28.4 & 24.4 & 2.2 & 102.4 & 26.8 & 1.6 & -0.0 & 0.0 & \\
8   &  1.29 $\pm0.05$  &  1.28 $\pm0.01$ &  +0.01 $\pm0.05$  & 43.7 & 27.0 & 1.1 & 2.9 & 44.0 & 18.7 & 1.8 & 80.9 & 42.1 & 1.3 & -0.3 & -0.2 & Shell 1 of $\dagger$: $(V-R)_J=0.68 \pm0.18$\\
10  &  1.40 $\pm0.13$  &  1.31 $\pm0.03$ &  +0.09 $\pm0.13$  & 55.0 & 17.5 & 0.6 & 0.9 & 55.0 & 9.7 & 0.5 & 11.8 & 54.2 & 0.7 & -0.2 & -0.1 & \\
13  &  1.40 $\pm0.07$  &  1.31 $\pm0.03$ &  +0.09 $\pm0.08$  & 66.6 & 21.8 & 1.4 & 0.6 & 67.3 & 17.8 & 0.4 & 26.9 & 65.1 & 1.4 & -0.1 & -0.2 & \\
15  &  1.22 $\pm0.14$  &  1.29 $\pm0.02$ &  -0.07 $\pm0.14$  & 78.1 & 12.5 & 1.7 & 0.4 & 79.0 & 14.2 & 0.3 & 17.7 & 77.6 & 1.3 & 0.1 & -0.1 & \\
17 & 1.23 $\pm0.15$ & 1.31 $\pm0.02$& -0.08 $\pm0.15$  &                &  & & &  & &  & &  & &  & &\\
\hline
\object{N 5982}&   &     &           &        &      & &   &  & &  & &  & &                                   & &      P.A.=107.0 and ellipticity=0.31\\
\hline
\object{N 7626}&   &     &            &       &      & &   &  & &  & &  & &                                   & &      P.A.=15.0 and ellipticity=0.15\\
\hline
2 &  1.46 $\pm0.04$    &  1.29 $\pm0.02$ &  +0.17 $\pm0.04$ & 32.2 & 28.6 & 0.8 & 6.0 & 32.4 & 19.0 & 2.8 & 106.9 & 30.9 & 1.1 & -0.3 & -0.1 & $(B-V)_{shell}=1.11\pm0.05$\\
3  &  1.24 $\pm0.07$    &  1.28 $\pm0.02$ &  -0.04 $\pm0.06$ & 42.7 & 42.2 & 2.3 & 1.0 & 43.4 & 25.0 & 0.8 & 76.4 & 39.6 & 2.8 & -0.4 & -0.2 & \\
\hline

\label{shells}
\end{tabular}
\caption{Shell properties. (1) galaxy NGC number/shell indexation (see Figures A.1-F.1); (2) V-I colour and errors of the shells, (3) local V-I colour and errors of galaxy; (4) Colour difference between shell and local galaxy (5-8) fitted parameters of Gaussian model (see Section 3.6): $R_0$, $\Phi$, $\rho$ and $R_g$; (9-11) fitted parameters of $(r_0-r)^{-0.5}$ model (see Section 3.6): $R_0$, $\Phi$ and $\rho$; (12-16) fitted parameters using Gaussian-Hermite fitting:  $\gamma$, $R_0$, $\sigma$, $h_3$ and $h_4$; (17) Comments: Average ellipticity 1-$\frac{b}{a}$ and position angle of the outer parts of the galaxy; comparison with external data with $\dagger$ = Fort et al. (1986)}
\end{sidewaystable*}

\subsection{NGC 1344}
\label{sec:ngc1344}
The shells in this galaxy are supposed to be the result of phase wrapping, since NGC 1344 is a type I shell galaxy.
The colour of one outer shell of this type I shell galaxy was determined by \citet{carter82}. This shell appears to be somewhat bluer than the main body of the galaxy. 
In our GALPHOT residual data the innermost shell is visible at about 27'' North from the centre. For the first time, we determined positions and the colours of some inner shells for this galaxy. The shell positions (Table 2) seem to show an interleaving pattern, although not as regularly as in NGC 3923. The V-I colours of the Northern shells are redder than the galaxy while they are similar or slightly bluer than the galaxy on the South side. The redder colours on the North side are probably due to dust, since these shells are much nearer towards the centre where dust is more present. The blue colour of one of the Southern shells is at least consistent with the earlier finding of Carter et al. The low number of shells and blue colour of some of these may be evidence for a relatively recent merger event. 




\subsection{NGC 2865}
\label{sec:ngc2865}

This galaxy is classified as a type II shell galaxy. However, looking at our residual images this system looks more as if it belongs to the type III class, with lots of irregular features and loops. The core is much bluer (V-I $\approx$ 1.05) than the outer parts (V-I $\approx$ 1.15). This blue colour is related a young stellar population (0.4-1.7 Gyr) which forms a KDC \citep{hau}. NGC 2865 also contains an incomplete HI disk \citep{shimi}. An interesting result is that a bright HI patch is coinciding with a bright shell observed earlier (labelled as shell 2 in this work and called shell 2B by \citet{fort}). Figure 1c of Schiminovich et al. shows an overlay of the HI data and Fort's schematic shell map. If these are related, this means that the shell is moving towards us, implying a spatially wrapped shell and confirming that this is not a type I shell galaxy. However, the asymmetric shell profile seems more consistent with a phase wrapped shell.
Fort et al. derived a slightly redder colour for shell 2 of $(V-R)_{Johnson}$= 0.84 $\pm0.09$, compared to the galaxy colour $(V-R)_{J}$=0.75 $\pm0.03$. A similar colour difference shell 2 and the galaxy is found in this work: V-I=1.11 $\pm0.05$ vs local galaxy V-I=0.98 $\pm0.01$. The 'jet', as reported by Fort et al. looks more like a loop (see top of Figure C.1), which may or may not be connected with the bright shell and HI.

Compared with the simulations, the residual image looks similar to situations in for instance Figure 2 or 6 of \citet{h0}, Figure 10 of \citet{h2}, or Figure 11 of \citet{dup}. All these simulations use small disk galaxies as the intruders, which is supported by the presence of an HI disk in NGC 2865.


The bright shell 2, coinciding with the HI patch, is probably the best candidate for follow-up spectroscopy of our whole sample, because of its high contrast with respect to the galaxy light ($\approx$0.8 mag. higher in V than the galaxy, see Figure 4). 


\begin{figure}
    \resizebox{\hsize}{!}{\includegraphics{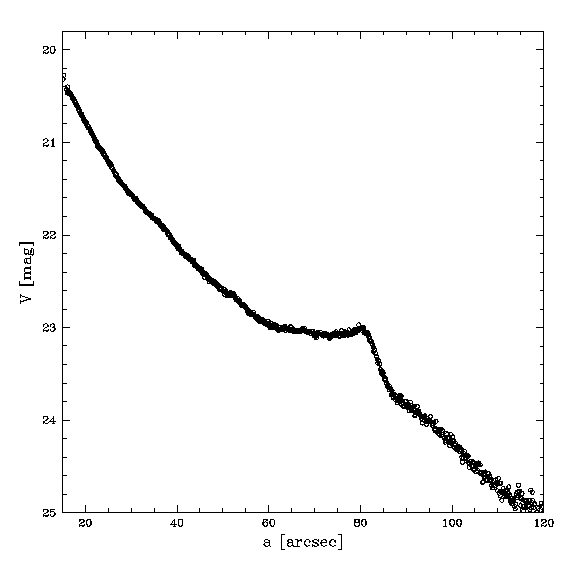}}
    \caption{Light profile of the galaxy along the wedge containing the brightest shell 2 of NGC 2865. The shell is about 0.8 magnitude brighter in V than the galaxy.}
    \label{2865shell}
\end{figure}


\subsection{NGC 3923}
\label{sec:ngc3923}

This is probably the most studied shell galaxy and the prototype of the Type I shell class, where shells are placed in interleaving order along the major axis. \citet{prieur}, using ground based CCD data and photographic plates, mapped the whole shell system. A comparison shows that we do not find any other shells than those already given by Prieur, despite our much better resolution and galaxy subtraction near the centre (compare our Figure D.1 with his Figures 1-5). In the last column of Table 2, we list and compare his shell positions with our results. There is a good agreement; small offsets between positions are likely due to measurement errors.
Fort et al. (1986) give colours for three shells in NGC 3923, with only one of them in our field of view (our shell 8): his shell 1 has a colour of $(V-R)_{Johnson}$ =0.68 $\pm0.18$ with the local galaxy colour being about $(V-R)_J=0.82 \pm0.03$. Our results for this shell are V-I=1.29 $\pm0.05$ and local galaxy colour of V-I=1.28; the colour difference being consistent with Fort' s work. On the West side, we find that all shells have a similar or slightly redder colour than the galaxy. We were not able to determine reliable shell colours on the East side. The local models, using different degrees for the Legendre polynomials, do not give stable answers. This is related to the low S/N of the images and badly defined shell features.





\subsection{NGC 5982}
\label{sec:ngc5982}

NGC 5982 is a Type I shell galaxy (Figure 5). The galaxy is well known for its KDC \citep{wagner}, recently confirmed using 2D central mapping using OASIS \citep{mcdermid} and SAURON \citep{emsellem}. Analogy with the KDCs of Emsellem et al. (2004) indicates that the KDC is probably a rotating central feature, i.e. a disk. The ellipticity becomes very round in the inner 2'' of the galaxy. We were not able to smoothly fit the central regions. Even after using 4 harmonics, a quadrupole with wings containing a flux of about 200 counts ($\pm$ 1-2 \% of the galaxy flux), is still visible. The C4 coefficient in this region is about -0.02 (Figure E.3) indicating a boxy structure \citep{carter78}. 2D spectral mapping of this region \citep{emsellem} show that the stellar velocities exhibit a 90 degree offset in the central region with respect to the outer regions. 

The bad GALPHOT fit limits us in giving a final answer about how far the shells do extend to the centre. The innermost detectable shell is located about 8'' East. Assuming the galaxy is about two times further away than NGC 3923 (see Table 1), the distance from the centre of this shell is comparable to the inner shell of NGC 3923. Double checking using other methods (unsharp masking and GALFIT \citep{peng}, which fits symmetric 2D models), show no more inner shells. 

Unfortunately reliable shell brightness could not be determined, because of their faintness and low contrast. A deeper, 3600s, ground based image, was found in the ING Archive (taken with the 2.5 meter Isaac Newton Telescope INT in July 1989 in the R band). After again using GALPHOT and subtracting galaxy-model, we obtain Figure 5, which shows the shells with somewhat higher contrast. The outer shell 24 on the minor axis is barely visible in the ACS images and is 90 degrees displaced with respect to the inner shells . The next bright outer shells 20 and 21 are slightly displaced with respect to the inner shells. This peculiar shell morphology resembles the scenario shown in Figure 6c of \citet{dup}. In this simulation a small elliptical falls into the potential of a prolate E3.5 galaxy with an impact angle of 90 degrees. The same misalignments occur for the outer shells in this event.

\begin{figure}
    \includegraphics[width=9cm, angle=0]{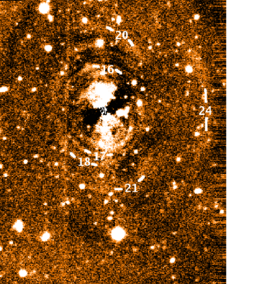}
    \caption{Ground based residual image of the shell system NGC 5982, with shell labels (see Table 2). This morphology resembles the scenario shown in Figure 6c of \citet{dup}, where a small galaxy has fallen in the host galaxy at an offset angle of 90 degrees.}
    \label{5982}
\end{figure}

This galaxy is classified as an YP galaxy (i.e. containing a young stellar population similar to NGC 2865) with a fine structure index $\Sigma_2$ of 6.8 \citep{michard}. Michard \& Prugniel do not mention shells. Looking at our images we estimate that there are at least eight shells, which would imply a much higher $\Sigma_2$ of about 11. 



\subsection{NGC 7626}
\label{sec:ngc7626}

Its core kinematics \citep{balcells} and the fact that it has bright globular clusters (Paper I) indicate a (minor) merger event which happened one or several Gyr ago. This is the first time that shell colours and brightness have been obtained for this galaxy.
A shell on the East side was detected by \citet{jedr2}; this shell lies outside our field of view. Another possible shell lying on the S.W. side, was detected by \citet{ft}. This shell is also seen in our data. We only detect one other, very sharp, edged shell to the N.E. The structure of these shells looks somewhat like those of the simulations shown in Figure 5 of \citet{dup}. Here a a small spiral (1\% mass of host) was thrown into the potential of an E3.5 oblate galaxy. Frame 3 (after 4 Gyr) of this Figure looks very similar to the NGC 7626 shell system. The zero rotational momentum encounter should create phase wrapped shells which is probably evidenced by the radial shell shapes (see Section 5.2). Further support for a disk intruder galaxy comes from the shell colours. In both (B-V) and (V-I), the brightest, inner shell is redder than the galaxy (see Figure 6), which is probably due to dust (see Section 5.3). The fainter outer shell has a similar or slightly bluer (V-I) colour than the galaxy.

\begin{figure}
    \resizebox{\hsize}{!}{\includegraphics{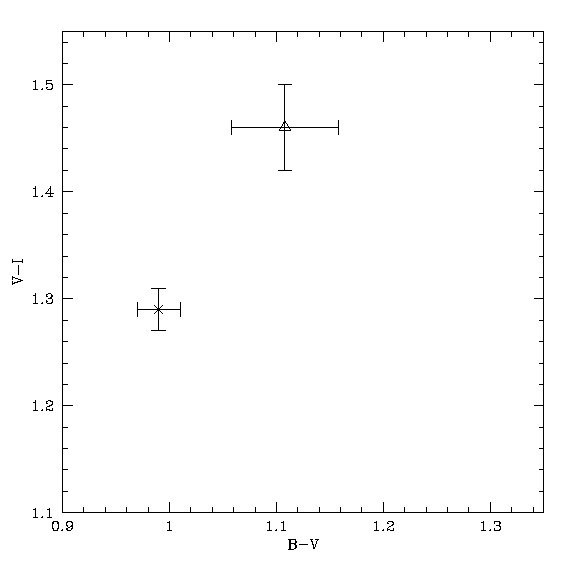}}
    \caption{Colour-colour, B-V vs V-I, diagram of the brightest, inner, shell in NGC 7626. Cross + errorbars represent the local galaxy colour, while the triangle with errorbars represents the derived shell colour}
    \label{7626shells}
\end{figure}


\section{General results}
\label{sec:results}

\subsection{Shell radial distributions}
\label{sec:results:shellradii}



As discussed in Sect.~\ref{sec:introduction}, inner shells contain useful clues to the shell formation process, and indeed, one of the central goals of the HST-based
imaging program was to determine how close to the galaxy centres we find shells. Our determined shells and shell radii (see Sect.~\ref{sec:analysis:shellradii}) are listed in Table 2, given in arcsec as well as in units of the effective radius \reff\ (from col.~14 of Table 1).  

Innermost shell radii span a wide radial range, from $\rmin/\reff=0.24$ in NGC~5982 to $\rmin/\reff = 2.86$ in NGC~2865.  We find a slight tendency for type-I shell systems to extend within \reff\ and for types II and III to lie in the outer parts: the two innermost shells, in units of \reff, are in NGC~5982 and 3923, two Type-I galaxies, while the third galaxy with $\rmin/\reff < 1$, NGC~474 (type-II), has an uncertain value of \reff: as shown in Sect.~\ref{sec:globalparameters}, \reff\ might be significantly smaller than listed in Table 1; adopting such smaller value for \reff, we would get $\rmin/\reff =4.2$.  However, not all type-I shell systems show inner shells: for NGC~1344 we find $\rmin/\reff>2$.  

The $HST$ imaging has revealed shells in the inner two kpc for two of the type I shell galaxies: NGC~3923: $\rmin = 8\arcsec = 1.8$~kpc (at an assumed distance of 20.0 Mpc) and  NGC~5982: $\rmin = 8\arcsec = 1.7$~kpc (at an assumed distance of 41.9 Mpc).  This is interesting as it clarifies that shells do indeed form near the galaxy nuclei.  This region is easily available to spectroscopic kinematic measurements, hence the correspondence between kinematic features and shells may be revealed.  

For the galaxies with no inner shell detections, the most straightforward interpretation is that shells never formed at those radii. This conclusion may be too simplistic.  
Three conditions need to concur for the detection of inner shells: ($i$) shells need to form; ($ii$) shells need to survive until the observation epoch; and, ($iii$) they need to be detectable through the shell-detection methods employed.

The second and third conditions listed above make shell detection harder as we look closer to the galaxy centres.  We first address shell survival. We expect shell lifetimes to be shorter near the centres: phase mixing scales with dynamical time, hence it is faster near the centre; shells should lose contrast and fade away faster near the centres.  Furthermore, galaxy centres are more dynamically active than the outer parts, as any object that merges after the formation of the shell system and reaches the centre, will gravitationally perturb the orbits of the stars that define the shells. In this respect, it is quite suprising that we see inner shells in NGC~3923 and NGC~5982, which have a large number of shells and are relatively old systems \citep{nulsen}

Shell detectability becomes progressively more difficult as we approach galaxy centres.  As mentioned above, the pronounced brightness gradients in the inner regions of ellipticals lead to the break-down of unsharp-mask techniques.  Because shells near the centres are closer to each other than further out, it may become difficult for the detection algorithm to pick up the underlying, non-shell brightness levels.  Finally, elliptical nuclei are known to be dusty \citep{lauer95,phillips96,peletier99,ravi}.  Dust may act in two ways.  It may simply \textsl{hide} the shells: examples of such an effect are NGC~474, NGC~1344 and NGC~7626.  Dust may also perturb the general light distribution, so that the smooth galaxy model one generates and subtracts to reveal the shells has too strong residuals for the shells to appear.  Typically, the underlying light distribution needs to be smooth to within a few percent for the shells to show up.  Strong dust patches easily lead to stronger third- or fourth-order Fourier residuals in the isophotes.  Examples of this situation are NGC~2865 and NGC~5982. Clearly, $HST$ imaging at NIR wavelengths would allow us to see through the dust and would strongly improve the chances of detecting inner shells in ellipticals.

\subsection{Shell brightness profiles}
\label{sec:results:shellprofiles}
Looking at Figure 3, it is clear that in general the Gaussian model fits the data better than the $(r_0-r)^{-0.5}$ model. The latter model especially fails for lower r. However, this model is only meant to describe the flux behaviour very close to the shell maximum at projected radius $r_{max}$. The $(r_0-r)^{-0.5}$  model also predicts a fast dropping flux at distances slightly larger than $r_{max}$. This is indeed seen at the two bright shells in NGC 7626 (see Figure 3). Here the flux drops from its maximum to zero within a small interval relative to the shell size. In general this seems to happen for shells with a plateau, i.e. N474\_6, N474\_13, N2865\_2, N3923\_8, N7626\_2, N7626\_3. 

The $(r_0-r)^{-0.5}$  model is supposed to describe phase wrapped shells \citep{dup,prieur}, but is does not seem to fit well most of the bright shells in type I shell galaxies NGC 1344 and NGC 3923, where shells are expected to be the result of phase wrapping. For NGC 3923 this may be related to the age of the shell system. The large number of shells indicate an old age. The shells may smooth out as a result of their older age and will not have razor sharp edges as the $(r_0-r)^{-0.5}$ model assumes. On the other hand: NGC 1344, showing only a few shells, indicating a younger age, does not do much better. NGC7626, which is probably very young (see Section 4.6), might be a better example where the model works. We conclude that the $(r_0-r)^{-0.5}$  model works best for younger shells. Older phase wrapping shells probably have a more extended structure and density profile than the theoretical $(r_0-r)^{-0.5}$ model, for instance due to internal velocity dispersions in the intruder galaxy. It seems that the $(r_0-r)^{-0.5}$ is just too simplistic to describe the real shells.

\subsection{Shell colours}
\label{sec:results:colours}

In our galaxy sample we find only one shell with blue colours. All other shells have similar or redder colours. Red shell colours are also found by many others like in NGC 7600 \citep{turn}, IC 1459 \citep{forbes2}, NGC 7010, NGC 7585 and IC 1575 \citep{pier}. The redder colours could be explained by at least four scenarios:
\begin{itemize}

\item The stars in the shells are on average older than those in the main body of the galaxy. This may occur if the shells belong to older parts of the intruder galaxy, e.g. the bulge, or if the interaction or merger led to the formation of young stars from gas throughout the galaxy. N-body simulations show that the best reproductions of shell morphologies are obtained by using very small intruder galaxies, with only a few percent of the mass of the host. It is therefore not very likely that they will form sufficiently many new stars which could lower the average colour of the whole galaxy.
\item The stars in the shells are more metal rich than those in the main body of the galaxy. This is highly unlikely given the expected small mass of the intruder galaxy and usually positive correlation between metallicity and galaxy mass \citep{sandage}.

\item The stars in the shells have different, redder, populations than the underlying galaxy. This scenario only works for very specific conditions. If the progenitor galaxy is a (small) late type (star-forming) galaxy and star formation is truncated after the merger event, then after some several $10^8$ yrs the light of the original stellar population will be dominated by RGB and AGB stars, which will redden the integrated colours of the shells. This reddening effect has been demonstrated by \citet{maraston}. Her Figure 27 (middle left panel) shows an enhanced reddening in V-I after several 100 millions of years, mainly due to AGB stars,. However, the amplitude of the reddening she found is not enough to account for our red shell colours, which are sometimes even redder than the elliptical galaxy colour. Resolved data on the shell or stream in M 83 also shows significant amounts of AGB and RGB stars \citep{dejong}, but his data are not deep enough to calculate a reliable global colour for this stream.

\item The shells contain more dust per unit stellar mass than the main body of the galaxy. Here, the problem is to explain why shells have more dust per unit stellar mass. Several possibilities can be thought of. The first possibility is related to the previous item: if RGB and AGB stars make up a significant part of the population, their large mass loss (mostly AGB stars) will result in more dust residing in the shells \citep{athey}. Another possibility is based on theory: it has been shown that it is possible for gas or dust to remain connected with the shell stars after a small merger \citep{kojima,charmandaris}. Observational evidence for significant amounts of dust residing in a shell was found in NGC 5128 \citep{stickel}. HI gas in shells has been found in M 83, NGC 2865 and NGC 3656 \citep{shimi, bal}. A third, speculative, idea explaining the presence dust in shells, is that dust is swept up by the shell stars as they pass through the potential of the galaxy. This should be tested using simulations. In the ISO archive, we found ISOCAM \citep{kessler} data for NGC 1344 and NGC 7626. These observations, taken at wavelengths near 9.5 $\mu m$, could in principle detect the warm dust. Although both galaxies show red shells, we find no evidence for enhanced emission at the shell regions in these two galaxies.
\end{itemize}





\subsection{Dust in the centres of shell galaxies}

All of our galaxies show visible dust features, mainly found in the central parts of the galaxy (see Figures 2 in Appendix A-F). Following \citet{tran} the morphologies of the visible dust features can be divided into two groups, i.e.: 1) nuclear ring or disks-like structures and 2) filamentary structures and/or (small) dust patches. All our galaxies show at least features of group 2 (filaments: NGC 474, NGC 2865, NGC 3923 and NGC 7626; small dust patches: NGC 1344, NGC 2865, NGC 3923 and NGC 5982). Although NGC 5982 was listed before as a dust-free galaxy \citep{sarzi}, we see several patches in the residual frames (two dust patches are visible in  Figure E.2 at about 6.5'' E and N.E.).

NGC 7626 shows both a dust lane and a nuclear ring within the inner arcsec (the ring was already reported by \citet{forbes}). A combination of these two dust features is not seen very often: Lauer et al. (2005) 
do not find any example in a sample of 77 early type galaxies. Saturation in the core in NGC 2865 and a bad fit of the inner regions in NGC 5982 hinders a conclusion about the presence of nuclear dust rings in these galaxies. The bad fit in NGC 5982 may be related to the presence of a KDC in the inner regions \citep{wagner,mcdermid}. When we apply GALPHOT to much lower resolution ground based data, we see similar (bad) results (see Figure 5). Lauer et al. are able to make a better fit, however they do not detect the dust patches and shells, probably due to smoothing in their modelling procedure. All dust morphologies are summarised in column two of Table 4. Column three of the same table lists the position angles valid for the filaments and nuclear disk. 

'Visible dust masses' were obtained by using the method of \citet{vand}. They assume that the visible dust acts as a foreground screen w.r.t. the background galaxy light, which will result in a lower limit for the dust mass and will have large errors of the order of 50\%. The dust mass is derived with the following expression:
\begin{equation}
M_d = <A_V> \Sigma \Gamma^{-1}
\end{equation}
with $<A_V>$ extinction measured at a pixel, $\Sigma$ the surface area (using distances listed in Table 1) and $\Gamma=6\times10^{-6}$ mag kpc$^2 M_\odot^{-1}$ the extinction coefficient per unit mass. Only areas were selected where the extinction/dust is visible by eye. $A_V$ was calculated for each visible dust feature by dividing the real and model images. Values of $A_V$ are all lower than 1. All derived dust masses are listed in column 5 of Table 4. The masses are of similar magnitude to those found by other authors (van Dokkum \& Franx, 1995; \citealp{tran}).


\begin{table}
\centering
\begin{tabular} {l r r r r r }
\hline
\hline

Galaxy&Morph.&$P.A._{dust}$&$P.A._{gal}$&Mass&$M_V$ \\
\hline
(1) &    (2)    &   (3)    &    (4)  &  (5)  & (6) \\ 
\hline
\object{NGC 474} & f   & 0    & 0   & 8.3 & -21.17  \\ 
\object{NGC 1344}& p   &      & 160 & 0.3   & -21.07\\ 
\object{NGC 2865}& f,p & 40   & 150 & 4.0  & -21.59 \\ 
\object{NGC 3923}& f,p & 45   & 49  & 3.9  & -21.92 \\ 
\object{NGC 5982}& p   &      & 105 & 0.3   & -21.91\\ 
\object{NGC 7626}& f,d & 135  & 0   & 2.6  & -22.16 \\ 
\hline
\label{dust}
\end{tabular}
\caption{Properties of dust. Column 2: Morphology of the dust where d=disk, f=filament, p=patchy dust; column 3 and 4: P.A. in degrees of main dust feature and galaxy respectively; column 5: dust mass in $10^3 M_\odot$ as determined using the method of \citet{vand}; column 6: absolute V magnitude of galaxy}
\end{table}

\subsection{Dust origin}
\label{sec:dustorigin}


Currently, there are at least four scenarios which explain the presence of dust in the centres of early type galaxies. The dust survival time, which depends on the main destruction mechanism (sputtering by the hot X-ray gas), is expected to be relatively low in the centres of early type galaxies ($10^7-10^8$ yrs depending on the electron density, \citep{draine,tielens}. A problem with these timescale calculations is that it does not take into account the effect of self-shielding in dust clouds. This may enhance the survival time considerably. The ubiquitous presence of dust in the centres of early type galaxies is difficult to explain without some rate of replenishment. \citet{math} showed that it is possible to form dust clouds in the centres of early type galaxies by accumulating dust from stellar winds. Other scenarios use external influences like accretion from flybys or mergers with other galaxies. We will now discuss the dust properties in the shell galaxy sample, where external influences are evident.

Using HST archival data, about half of all elliptical galaxies exhibit visible dust features, which are equally present in power-law and core galaxies (\citealp{lauer}: 47\% of 177 in field galaxies). Assuming that a similar dust detection rate of 50\% is representative for our galaxy sample and given the fact that dust is visible in all our galaxies, we can reject the statement that our sample belongs to the parent set of normal early type galaxies with visible dust at the 97.5\% level. However we should also take into account the possibility that our sample is biased by considering much higher dust prevalence in certain classes of early type galaxies. This happens for instance in radio-loud galaxies, having dust detection probability of about 90\% (van Dokkum \& Franx, 1995; \citealp{gijs3}). Our sample hosts one radio galaxy: NGC 7626 \citep{hibbard}. Second, we consider a possible bias due to the presence of ionized gas. It is well known that dust is almost always accompanied with ionized gas in early type galaxies \citep{mach,sarzi}. Due to selection effects in the detection of visible dust, the converse is not true, although the probability to detect dust in early type galaxies with ionized gas is still quite high \citep{tran}. Modern instrumentation detect emission in about 75\% of early type galaxies \citep{mach,sarzi}. Compared with these detection rates, our sample does not seem to be biased: only three galaxies (NGC 474, NGC 5982 and NGC 7626) have low levels of H$\alpha$+[NII] emission \citep{gijs3, gijs2} with luminosities below the median value of $2\times10^{39}$ ergs s$^{-1}$, determined for a nearly complete magnitude limited sample of nearby galaxies \citep{ho}. Combining these biases and dust detection probabilities still implies that we can reject the statement at the 95\% level, and that shell galaxies have a higher dust prevalence than normal early type galaxies, contrary to an earlier finding by \citet{sadl}. 


Number counts of the morphology of the dust features, occurring in our sample, also differs from normal early type galaxies. Dust features in the centres of early type galaxies come in two types: regular rings and irregular shaped patches or lanes. Number counts give a ratio of 3:5 for the two types of dust features. They are almost never both seen in one galaxy \citep{lauer}, which has been used as evidence for an (episodic) dust settling sequence scenario \citep{tran,lauer,gijs}. Here, first the irregular dust patches and lanes form in some way, while some time later these dust features move to the centre and form a disk in dynamic equilibrium. All the dust features in our sample seem to be out of dynamical equilibrium as they show up as irregular patches or lanes. The probability for this to happen, assuming a regular to irregular ratio dust feature of 3:5, is 6\%. These considerations lead to the conclusion that external influences are responsible for the ubiquitous presence of dust in shell galaxies. 


We will now briefly discuss the dust properties of the individual shell galaxies and see how they fit into this discussion and shell formation theory.

NGC 474 contains several pronounced dust lanes  within the inner 15'' (see Figure A.2), not detected previously (see e.g. \citealp{ravi} and \citealp{sarzi}). It contains the largest visible dust mass of our sample ($\approx 10^4 M_\odot$). Sarzi et al. report a misalignment of the stars and ionised gas by 74 $\pm 16$ degrees. Comparing the position angle of the dust lanes with the velocity maps of Sarzi et al. show that the dust is aligned with the velocity field of the ionised gas and not with the stars. This probably implies a connection between the dust and the gas and is seen in many other early type galaxies \citep{goud2,ferrari,sarzi}. The velocity structure of the gas, the dust morphology of NGC 474 (many dust lanes residing on top of the supposed bulge) and the large offset between stars and gas+dust disturb the picture of this galaxy being an S0, and point towards an external origin for the dust. 

The well defined type I shell galaxies, NGC 1344, NGC 3923 and NGC 5982, are expected to be created by a minor merger with a small, non-rotating dwarf galaxy. While the dust content of NGC 1344 and NGC 5982 is quite low (several time $10^2 M_\odot$, Table 4) and corresponds to such a scenario, the dust mass of NGC 3923 is an order of magnitude higher. Most (80\%) of the visible dust in NGC 3923 belongs to a large patch, visible in the NE direction at 50 arcsec from the centre (see Figure D.1). This patch was shown to be part of NGC 3923 and also emits small amounts of H$\alpha$+[NII] \citep{pence}. At the projected distance of this patch, the electron density \citep{fukazawa} corresponds to a dust sputtering minimum survival time of $4\times10^7$ yrs. Another 5\% resides in small patches at 4.5 arcsec from the centre (Figure D.2). At this projected distance, the dust sputtering minimum survival time is $\approx 10^6$ yrs. The rest, 15\%, is located in a long diffuse dust lane NW from centre. The amount of dust in this galaxy does not conform to the minor merger picture of a small elliptical dwarf galaxy falling into a much larger potential (already noted by \citealp{carter2}). An internal origin for the dust (e.g. ejection from stars) seems to be the most likely scenario. 

In NGC 2865 and NGC 7626, comparison of simulations and shell structures point towards gas rich intruder galaxies (see Section 4.3 and 4.6 respectively), which has resulted in recent (0.4-1.7 Gyr) star formation in the core of NGC 2865 \citep{hau} and likely the creation of new globular clusters (Paper I). The dust content of several times $10^3 M_\odot$ (see Table 4) is distributed in diffuse layers and patches near the centre (Figures C.2 and F.2). In NGC 2865 we can again calculate the minimum dust sputtering survival time because the electron density is known \citep{fukazawa}. This results between $3\times10^7$ and $1\times10^8$ years for the inner and outer dusty regions respectively. This is again much lower than merger timescale of several $10^8$ to $10^9$ yr.


\section {Summary}
Using observations in V and I with the ACS on board the HST, we analysed the properties of shell systems, in particular their colours, morphologies and dust properties. For most shells listed in this paper, we determined their colour for the first time. For those shells for which their colour had already been determined, we find similar results, giving support to the quality of the result of our methodology. In general we find that colours of shells are similar or redder to the colours of their host galaxies. We attribute the red colour to dust which is physically connected to the shell. In some cases, a different stellar population as a result from a truncation of star formation, may also redden the shells. The only blue shell is found in NGC 474, which is very likely related to a long tail and probably a very recent minor merger event. N-body merger simulations, rather than simulations by the interaction model, describe best the observed morphologies of the shell systems. 

We detect out of dynamical equilibrium central dust features in all our galaxies. Comparison with a set of 'normal' elliptical galaxies, which have a dust detection rate of 50\%, implies an external origin for central dust found in shell galaxies. However this is in contradiction with theoretical predicted dust survival times. Better models of dust behaviour in centres of early type galaxies, which include self shielding, are needed to solve this problem. The best shell candidate for follow-up spectroscopy has been found in NGC 2865.

Innermost shells are found in the type I shell galaxies NGC 3923 and NGC 5982 at a distance about  2 kpc from their centres. These shell have survived for a long time, since both galaxies have relatively old shell systems.

Current models to describe the profiles of phase wrapped shells probably work best for young shells.


\section {Acknowledgements}
This research is partially based on data from the ING Archive.

\listofobjects
\clearpage
\appendix
\clearpage
\section{Results for NGC 474}
\begin{figure}
    \includegraphics[width=9cm, angle=0]{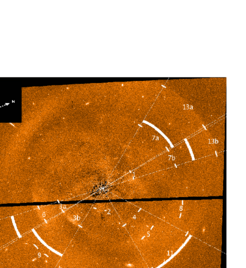}
    \caption{ACS residual image of GALPHOT for NGC 474 in V, surface brightness of shells were determined in the wedge regions using ellipticities as indicated by the white strips. The size of the ACS field of view is 202 x 202 arcseconds.}
    \label{res474}
\end{figure}

\begin{figure}
    \includegraphics[width=7.2cm, angle=0]{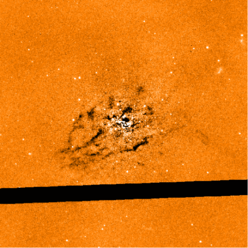}
    \caption{Inner region residuals of NGC 474 in V (40x40 arcsec). Dust is directed in the NS direction and aligned with the ionized gas (to be compared with the Figures in \citet{sarzi})}
    \label{inn474}
\end{figure}



\begin{figure}
    \includegraphics[width=9cm, angle=0]{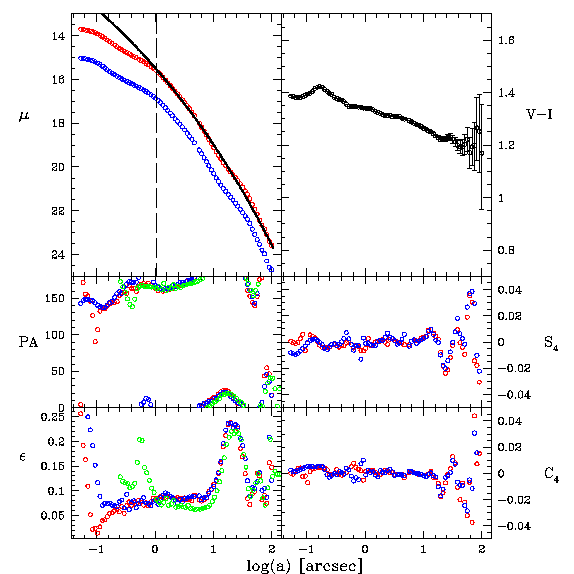}
    \caption{Morphological data NGC 474, blue and red curves represent V and I band data respectively. Top left: surface brightness profiles, corrected for background (see text). The black line is a S\'ersic fit to the I band surface brightness data right of the vertical dashed line. Middle  and bottom left: position angle and ellipticity respectively. Green data points represent WFI data. Top right: Global V-I profile. Top middle and bottom: S4 and C4 respectively.}
    \label{morf474}
\end{figure}

\begin{figure}
    \includegraphics[height=9cm,width=9cm, angle=0]{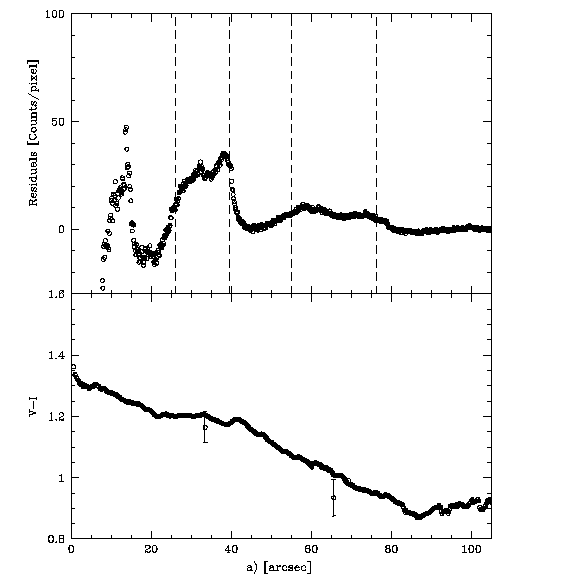}
    \caption{Photometric data in the wedge covering shells 3b and 9. Top panel: V band flux in counts averaged along wedge using ellipticities as indicated by the white strips drawn into the wedges in Figure A.1. The region within the vertical dashed lines was used to calculate the shell V-I colours which are shown in the bottom panel. The bottom panel also shows the galaxy V-I colour profile, calculated from a Voronoi binned image.}
    \label{Left0474}
\end{figure}

\begin{figure}
    \includegraphics[height=9cm,width=9cm, angle=0]{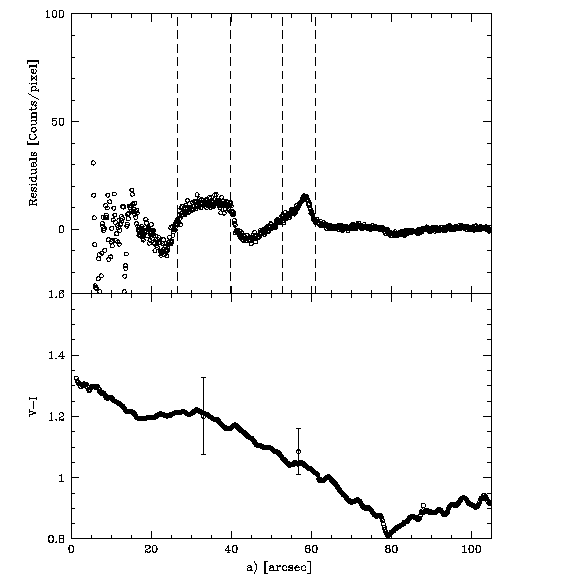}
    \caption{Photometric data in the wedge covering shells 3a and 6; Description: see Figure A.4}
    \label{Left1474}
\end{figure}

\begin{figure}
    \includegraphics[height=9cm,width=9cm, angle=0]{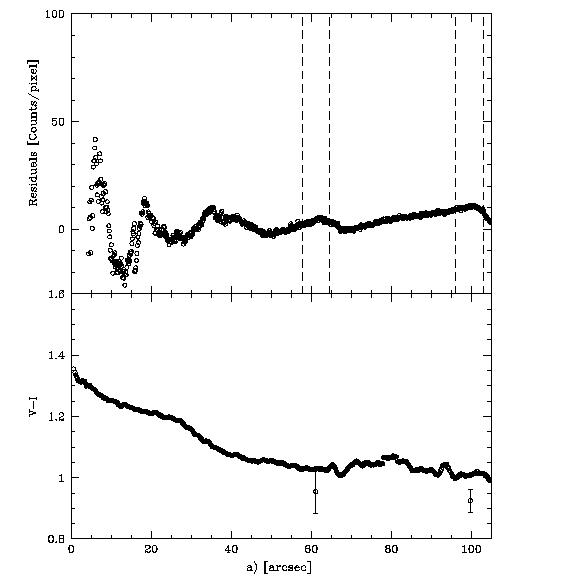}
    \caption{Photometric data in the wedge covering shells 7a and 13a; description: see Figure A.4}
    \label{Right0474}
\end{figure}

\begin{figure}
    \includegraphics[height=9cm,width=9cm, angle=0]{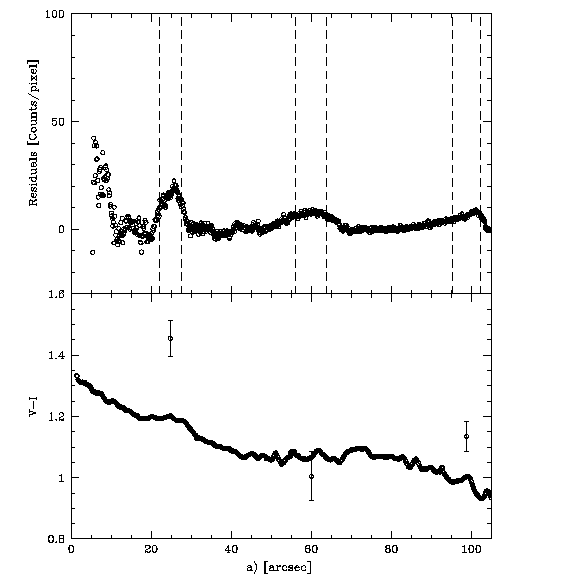}
    \caption{Photometric data in the wedge covering shells 1, 7b and 13b; description: see Figure A.4}
    \label{Right1474}
\end{figure}

\begin{figure}
    \includegraphics[height=9cm,width=9cm, angle=0]{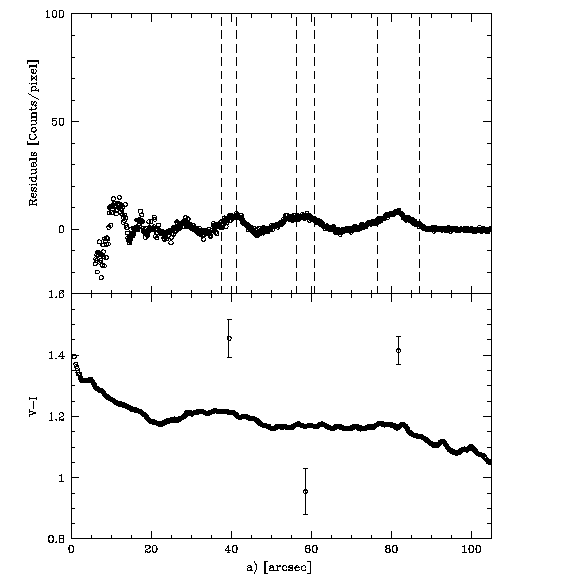}
    \caption{Photometric data in the wedge covering shells 4, 5 and 11; description: see Figure A.4}
    \label{Right2474}
\end{figure}

\clearpage
\section{Results for NGC 1344}

\begin{figure}
    \includegraphics[width=9cm, angle=0]{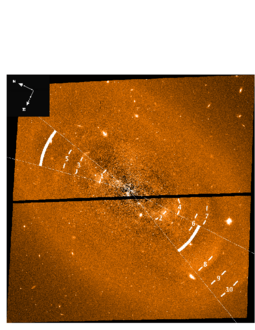}
    \caption{Residual image of GALPHOT for NGC 1344 in V with wedges left and right. The field of view is 202 x 202 arcseconds.}
    \label{res1344}
\end{figure}

\begin{figure}
    \includegraphics[width=9cm, angle=0]{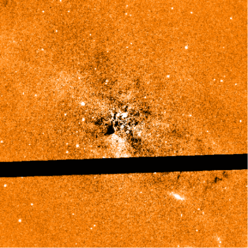}
    \caption{Inner region residual image of NGC 1344 in V (40x40 arcsec). Dust patches are visible.}
    \label{inn1344}
\end{figure}

\begin{figure}
    \includegraphics[width=9cm, angle=0]{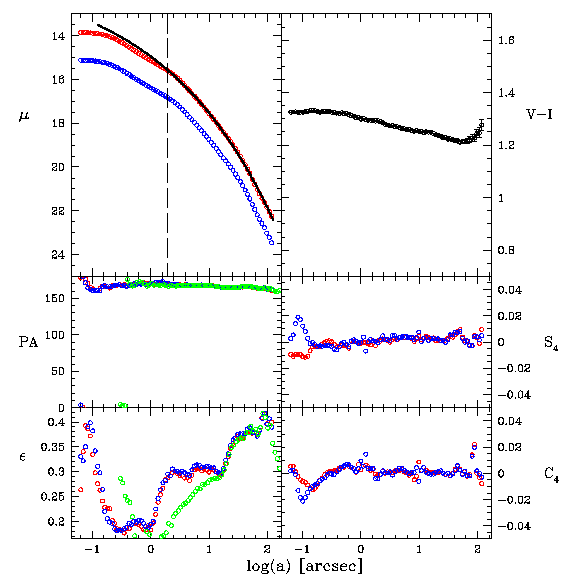}
    \caption{Morphological data NGC 1344. Description: see NGC 474, Figure A.3}
    \label{morf1344}
\end{figure}

\begin{figure}
    \includegraphics[height=9cm,width=9cm, angle=0]{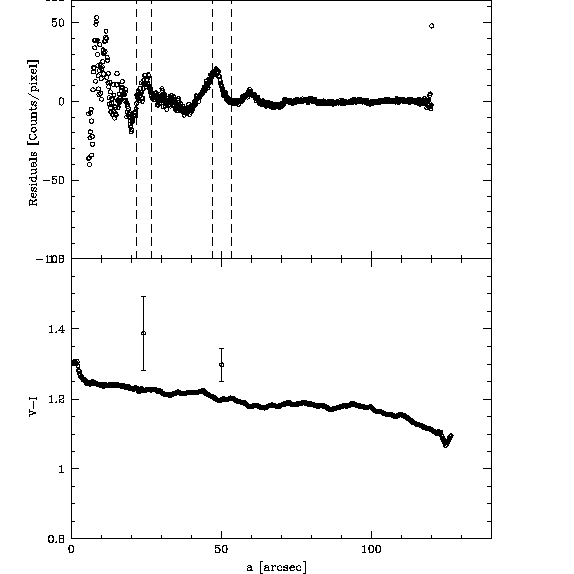}
    \caption{NGC 1344 shells 1 and 3 in the wedge on the North side. Description: see Figure A.4}
    \label{shell1344left}
\end{figure}

\begin{figure}
    \includegraphics[height=9cm,width=9cm, angle=0]{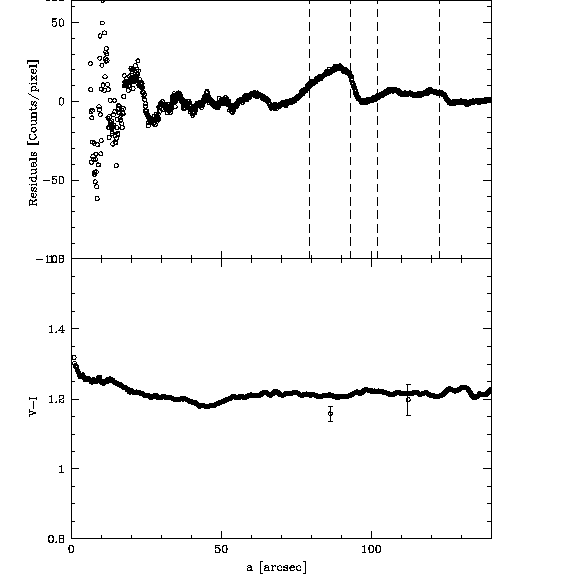}
    \caption{NGC 1344 shells 8 and 10 in the wedge on the South side. Description: see Figure A.4}
    \label{shell1344right}
\end{figure}

\clearpage
\section{Results for NGC 2865.}
\begin{figure}
    \includegraphics[width=9cm, angle=0]{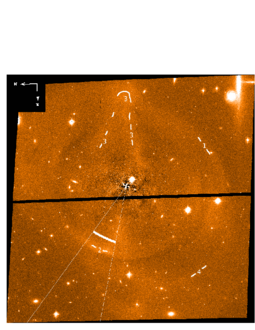}
    \caption{Residual image of GALPHOT for NGC 2865 in V with a wedge overlapping a bright shell. A loop is visible in the NS direction. This morphology was already drawn in \citet{fort}. The field of view is 202'' x 202''.}
    \label{res2865}
\end{figure}

\begin{figure}
    \includegraphics[width=9cm, angle=0]{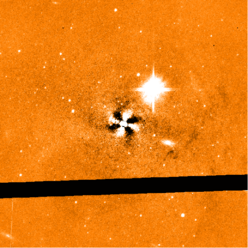}
    \caption{Inner region residuals of NGC 2865 in V (40x40 arcsec). Dust is visible in the South direction. Fitting problems in the central region are due to saturated pixels.}
    \label{inn2865}
\end{figure}

\begin{figure}
    \includegraphics[width=9cm, angle=0]{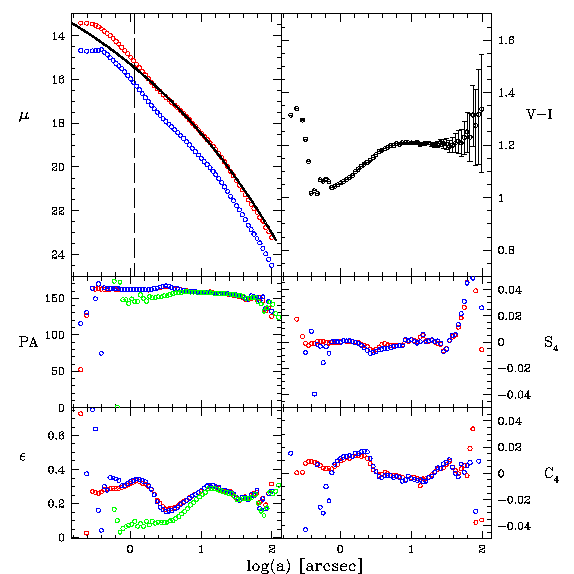}
    \caption{Morphological data NGC 2865. Description: see NGC 474, Figure A.3}
    \label{morf2865}
\end{figure}

\begin{figure}
    \includegraphics[height=9cm,width=9cm, angle=0]{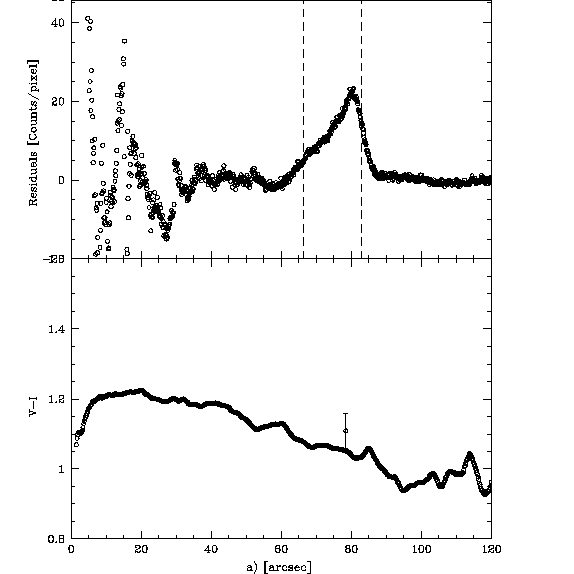}
    \caption{NGC 2865 shell 3 in the wedge placed at the East side. Description: see NGC 474, Figure A.4}
    \label{shell2865}
\end{figure}

\clearpage
\section{Results for NGC 3923}
\begin{figure}
    \includegraphics[width=9cm, angle=0]{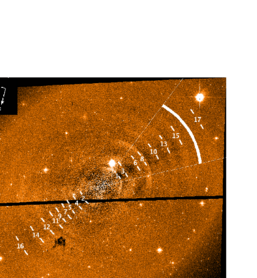}
    \caption{Residual image of GALPHOT for NGC 3923 in V with wedge. Note the large dust patch at the NE and faint dust lane within the wedge in the SW direction. The field of view is 202'' x 202''.}
    \label{res3923}
\end{figure}

\begin{figure}
    \includegraphics[width=9cm, angle=0]{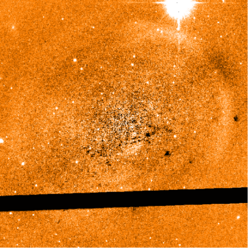}
    \caption{Inner region residuals of NGC 3923 in V (40x40 arcsec). Several small dust patches are visible. The innermost shell visible was also detected by \citet{prieur} using ground based data. No more other inner shells are detected.}
    \label{inn3923}
\end{figure}

\begin{figure}
    \includegraphics[width=9cm, angle=0]{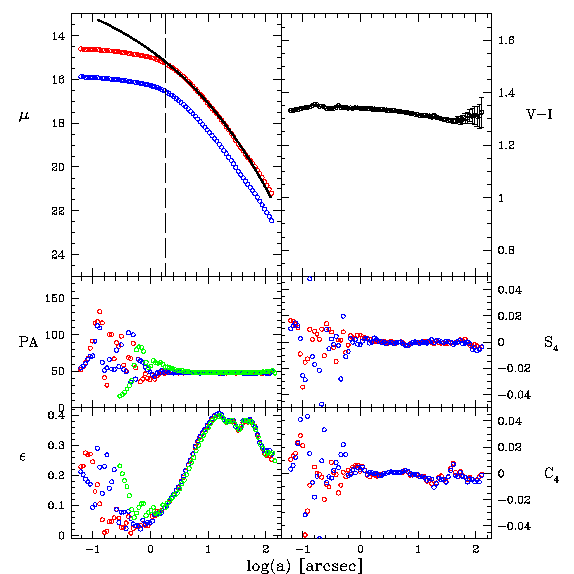}
    \caption{Morphological data NGC 3923. Description: see NGC 474, Figure A.3}
    \label{morf3923}
\end{figure}

\begin{figure}
    \includegraphics[height=9cm,width=9cm, angle=0]{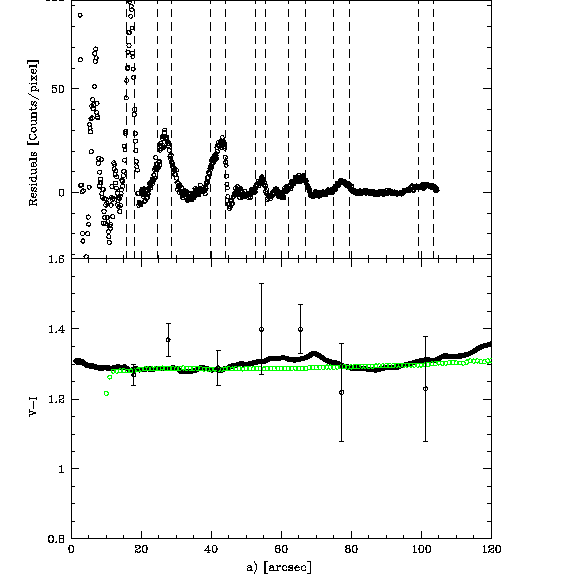}
    \caption{NGC 3923 shells 1, 3, 8, 10, 13, 15 and 17 in southern wedge. The lower panel shows shell and local galaxy V-I colours determined within the wedge: the points with errorbars are the shell colours. The solid line and green open circles represent V-I local colours from ACS and ground based VLT-FORS2 data respectively. The signature of the ring (the bump between r=40'' and 60'') is not visible in ground based data.}
    \label{shell3923}
\end{figure}

\clearpage
\section{Results for NGC 5982}

\begin{figure}
    \includegraphics[width=9cm, angle=0]{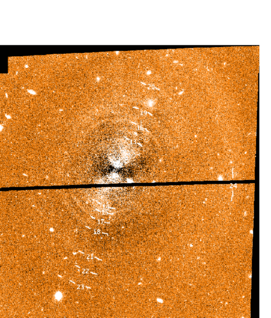}
    \caption{Residual image of GALPHOT for NGC 5982 in V. Shells are barely visible. The field of view is 202 x 202 arcseconds. The shell labels are best visible in the electronic version.}
    \label{res5982}
\end{figure}

\begin{figure}
    \includegraphics[width=9cm, angle=0]{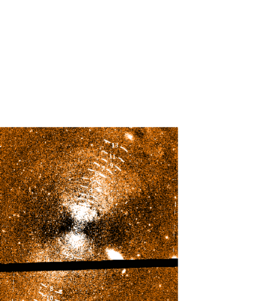}
    \caption{Inner residuals of NGC 5982 in V (60x60 arcsec). A small dust lane is visible on the major axis in the E direction, a more pronounced small dust patch is visible NE. The shell labels are best visible in the electronic version.}
    \label{inn5982}
\end{figure}

\begin{figure}
    \includegraphics[width=9cm, angle=0]{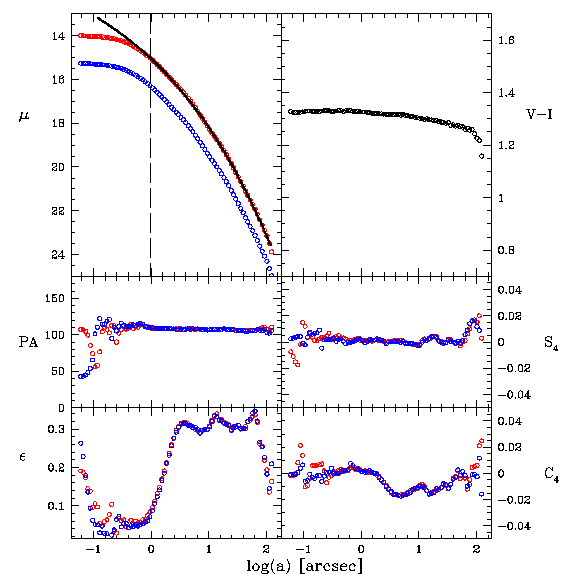}
    \caption{Morphological data NGC 5982.. Description: see NGC 474, Figure A.3}
    \label{morf5982}
\end{figure}

\clearpage
\section{Results for NGC 7626}

\begin{figure}
    \includegraphics[width=9cm, angle=0]{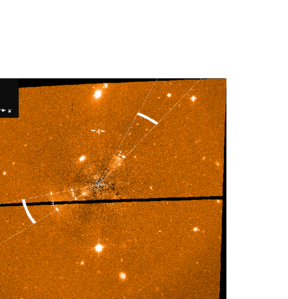}
    \caption{Residual image of GALPHOT for NGC 7626 in V . Two bright shells are visible with the wedges overlapping them. The field of view is 202 x 202 arcseconds.}
    \label{res7626}
\end{figure}

\begin{figure}
    \includegraphics[width=9cm, angle=0]{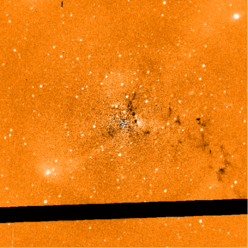}
    \caption{Inner region residuals of NGC 7626 in V (40x40 arcsec); the two shells are already outside the field of view, but the dust and a lot of globular clusters are clearly visible.}
    \label{inn7626}
\end{figure}

\begin{figure}
    \includegraphics[width=9cm, angle=0]{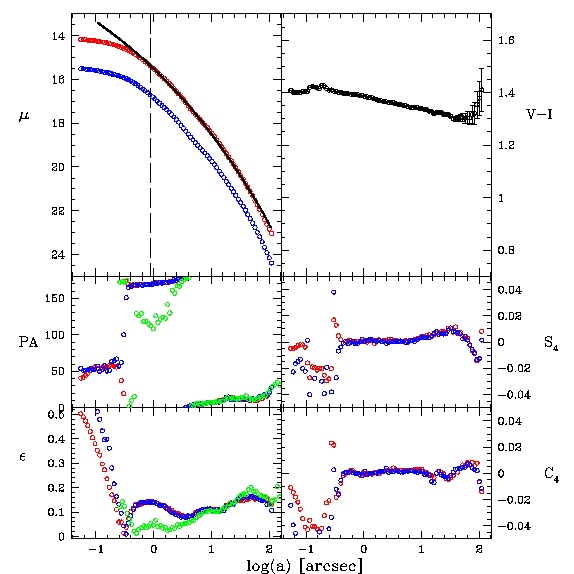}
    \caption{Morphological data NGC 7626.. Description: see NGC 474, Figure A.3}
    \label{morf7626}
\end{figure}

\begin{figure}
    \resizebox{\hsize}{!}{\includegraphics{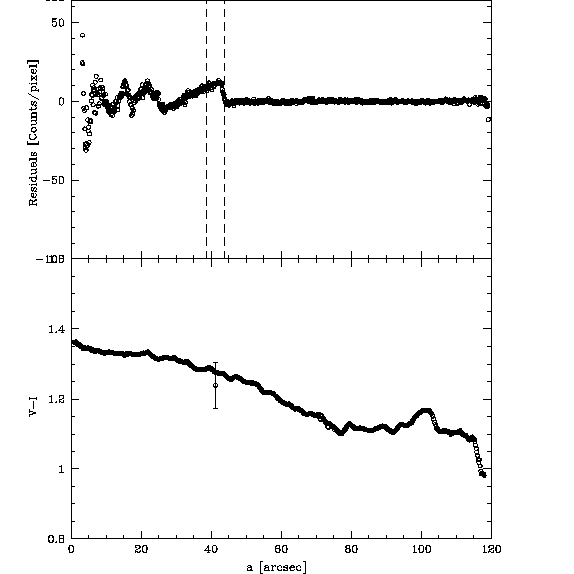}}
    \caption{NGC 7626 shell 3 in the wedge at the SW side. Description: see Figure A.4}
    \label{shell7626left}
\end{figure}

\begin{figure}
    \resizebox{\hsize}{!}{\includegraphics{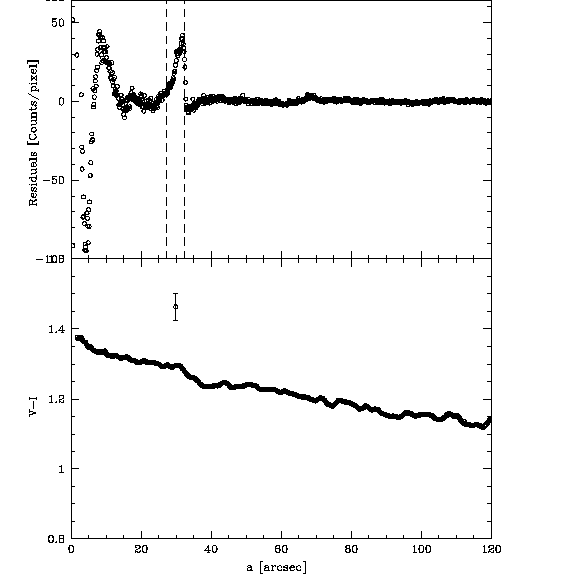}}
    \caption{NGC 7626 shell 2 in the wedge at the NE side. Description: see Figure A.4}
    \label{shell7626right}
\end{figure}

\begin{figure}
    \resizebox{\hsize}{!}{\includegraphics{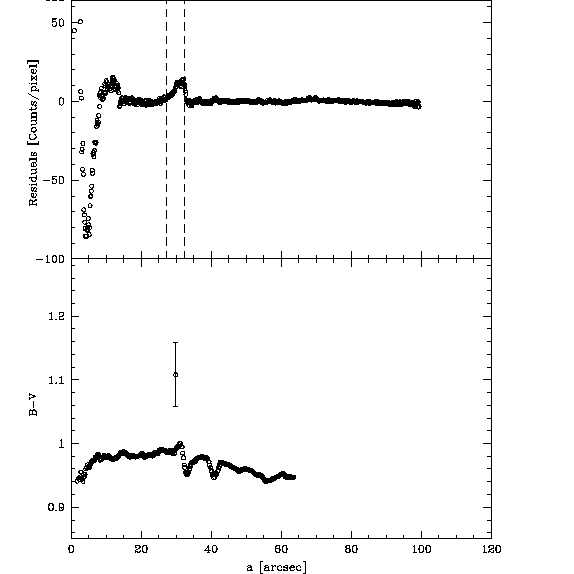}}
    \caption{Similar to previous Figure but now the B band residual flux is drawn in the top panel while B-V colours are drawn in the bottom panel.}
    \label{shell7626rightBV}
\end{figure}

\listofobjects
\end{document}